\documentclass[sn-basic]{sn-jnl}

\usepackage{graphicx}%
\usepackage{multirow}%
\usepackage{amsmath,amssymb,amsfonts}%
\usepackage{amsthm}%
\usepackage{mathrsfs}%
\usepackage[title]{appendix}%
\usepackage{xcolor}%
\usepackage{textcomp}%
\usepackage{manyfoot}%
\usepackage{booktabs}%
\usepackage{algorithm}%
\usepackage{algorithmicx}%
\usepackage{algpseudocode}%
\usepackage{listings}%
\usepackage{amsmath}%
\usepackage{natbib}

\jyear{2021}%

\theoremstyle{thmstyleone}%
\theoremstyle{thmstyletwo}%

\theoremstyle{thmstylethree}%

\usepackage{soul}

\begin{document}

\title[Article Title]{Understanding the formation of Saturn's regular moons in the context of giant planet moons formation scenarios}


\author[1,2]{\fnm{Michel} \sur{Blanc}}\email{michel.blanc@irap.omp.eu}

\author[3]{\fnm{Aur\'elien} \sur{Crida}}\email{aurelien.crida@oca.eu}

\author*[4,5]{\fnm{Yuhito} \sur{Shibaike}}\email{yuhito.shibaike@nao.ac.jp}

\author[6]{\fnm{Sebastien} \sur{Charnoz}}\email{charnoz@ipgp.fr}
\author[7]{\fnm{Maryame} \sur{El Moutamid}}\email{maryame@astro.cornell.edu}
\author[8]{\fnm{Paul} \sur{Estrada}}\email{paul.r.estrada@nasa.gov}
\author[2]{\fnm{Olivier} \sur{Mousis}}\email{olivier.mousis@lam.fr}
\author[9]{\fnm{Julien} \sur{Salmon}}\email{julien@boulder.swri.edu}
\author[2]{\fnm{Antoine} \sur{Schneeberger}}\email{antoine.schneeberger@lam.fr}
\author[2]{\fnm{Pierre} \sur{Vernazza}}\email{pierre.vernazza@lam.fr}

\affil[1]{\orgname{Research Institute in Astrophysics and Planetology, France}}

\affil[2]{\orgname{Laboratoire d'Astrophysique de Marseille, France}}

\affil[3]{\orgname{Université C\^ote d'Azur, Observatoire de la C\^ote d'Azur,
France}}

\affil[4]{\orgname{National Astronomical Observatory of Japan, Japan}}

\affil[5]{\orgname{University of Bern, Switzerland}}

\affil[6]{\orgname{Institut de Physique du Globe de Paris, France}}

\affil[7]{\orgdiv{Department of Astronomy}, \orgname{Cornell University, USA}}

\affil[8]{\orgname{NASA Ames Research Center, USA}}

\affil[9]{\orgdiv{IMCCE}, \orgname{Observatoire de Paris, France}}


\abstract{This article explores the different formation scenarios of the Kronian moons system in the context of a highly dissipative Saturn, with the objective of identifying the most likely of these scenarios. First, we review the diversity of objects - moons and rings - orbiting solar system giant planets, and the diversity of their architectures, which formation scenarios must reproduce. We then identify in this broader context the specific features of the Saturn system, such as the particularly large spectrum of its moon masses, the uniqueness of Titan and the presence of both dense and tenuous rings, before discussing the applicability of the different giant planet moon formation scenarios to the Saturn case. We discuss each of the most relevant scenarios and their respective merits. Finally, we tentatively propose a ``favorite'' scenario and we identify the key observations to be made by future space missions and/or Earth-based telescopic observations to validate this scenario or possibly alternative ones.}




\maketitle

\section{Introduction and objective of this chapter}\label{introduction}
It is generally accepted that giant planet moons are a by-product of the formation of their parent planets, similar in a way to our understanding of planets as a by-product of the formation of their parent star. Beyond this general statement, the different theories that explain the formation of giant planet moons differ by the way they describe the relationships and relative timing between the formation of the planet in the Solar Nebula (SN), the formation of its circumplanetary disk (CPD), the capture or accretion of satellitesimals in the CPD, the generation of rings, and the capture of planetesimals from the SN or, after its dissipation, from solid body reservoirs of the early Solar System.

This diversity of formation processes is not really surprising, knowing the large differences existing in the physical, chemical and dynamical characteristics of moons, as well as in their arrangement in space – the moon system architectures – among the only four giant planet systems of the Solar System. Among them, the Saturn system offers the broadest diversity of objects: in increasing order or radial distance, dense rings and tenuous rings, small moons inside or near the rings that are usually called "ring-moons", small and medium size regular moons outside the rings region, one large moon (Titan), and finally a large number of irregular moons. Thanks to the Cassini-Huygens mission, this system is also the one that has been best documented by close-in spacecraft observations that covered all families of objects, from small inner moons embedded in the rings to large irregular ones.

Attempting to trace back from these detailed observations of the contemporary Saturnian system the mechanisms and series of events that led to its assembly is therefore particularly relevant, because of the significant number of observational constraints that can be compared to model predictions. The main purpose of this chapter is to address this challenge by confronting predictions of these different formation models to observations. Since the Saturnian system is only one of the four giant planet systems of the Solar System, while the diverse formation mechanisms proposed may apply to different extents to all four of them, we place our study of the formation scenario of Saturn’s moons in the broader context of the diversity of formation scenarios proposed for the four giant planet systems.

This article is organized to follow this logic.

Section 2 is devoted to providing the broader context of the diversity of moons and moon system architectures.

We first review the variety of observational characteristics of individual moons and of moon systems architectures in the light of past and current Earth-based telescopic and space mission observations. On this basis, we identify the specific features of the different categories of Saturnian moons that have to be explained by candidate formation scenarios.

In section 3, we focus on the formation of the Saturnian moons system.

As an introduction, we review the different types of moon formation scenarios that have been proposed for the four giant planet systems: result of a giant impact; formation by accretion in a circum-planetary disk (CPD); formation by viscous expansion from a massive disk; capture… For each category of formation scenario, we discuss its relevance to the formation of the different giant planet systems and draw some conclusions on the key characteristics that each type of formation scenario tends to reproduce best.

Then we analyze the relevance of these different formation scenarios to the different categories of Saturnian moons and to the moon system as a whole. We discuss the merits of the different formation scenarios on the basis of available observations and in the context of the fast migration of moons which is the main subject of this Topical Series.

In Section 4, we draw some tentative conclusions on the relative success of the different models in reproducing the different categories of moons and tentatively propose a ``favorite'' scenario. We identify additional observations, laboratory data and simulation studies that are needed, first to discriminate among the different scenarios, and then to elaborate a more comprehensive scenario that will make it possible to assemble all the pieces of the moons formation puzzle.

\section{Characteristics of giant planet moons and architectures of their systems}\label{characteristics}

\subsection{Introduction}\label{intro2}
The moon systems of the four giants can be divided into three families which generally occupy three domains of radial distance. From inside out:
\begin{itemize}
    \item A first family of small, prograde inner moons, generally in close dynamical coupling to their planet’s rings and occupying near-circular orbits in the equatorial plane, are generally referred to as "ring-moons". They mainly reside within the Roche limit for ice, or extend only slightly beyond. Their sizes vary from a km to a hundred km at most;
    \item A second family of small-to-large-size regular prograde moons resides on near-circular orbits in the equatorial plane at distances of a few to a few tens of planetary radii. They are mostly decoupled from the rings, though a few of these objects display weak resonances with them. Their sizes range from a few 100 to a few 1000 km;
    \item Finally, a third family of small irregular moons is found at all four giant planet systems on mid-to-high-inclination, eccentric orbits extending to radial distances encompassing a significant fraction of the Hill radius of their parent planet. A majority of the smaller size component of this population occupy retrograde orbits, while a smaller number of larger size objects, above about 50 km diameter, dominate on prograde orbits. 
\end{itemize}
We successively describe the observational characterizing of these three families, ending up with the regular moons which are the main focus of this article.

\subsection{Small inner moon systems coupled to the rings}\label{small}

\begin{figure}[htbp]%
\centering
\includegraphics[width=0.9\textwidth]{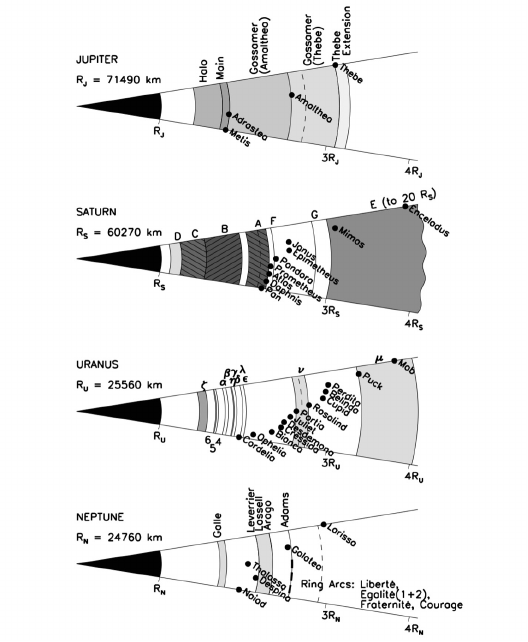}
\caption{Schematic comparison of the ring-moon systems of giant planets. The Roche radius for ice is shown as a dashed line. From \citet{bla21}.}
\label{fig:ring-moon}
\end{figure}

Figure \ref{fig:ring-moon} illustrates the intricate relative geometry of the radial distributions of small inner moons and rings in the four giant planet systems, relative to their planet's radius. Moons are indicated as black dots while the different rings are represented as gray areas, with increasing intensities of gray representing increasing densities of ring components.

Direct inspection of this figure reveals the extreme diversity of these small moons and of their relation to rings at the four planets.

Jupiter hosts only four small regular moons, i.e. in order of increasing planetocentric distance Metis and Adrastea (on very close orbits), then Amalthea and Thebe. The orbits of these moons coincide with sharp boundaries in Jupiter's tenuous dust rings: the main ring for Metis and Adrastea, and the Gossamer rings for Amalthea and Thebe. This has been interpreted as a result of the generation of an extended dust ring along these moons orbits by meteorite impact on their surfaces, which subsequently drifts inward under the combined effects of dust charging and motion in the Jovian magnetic and electric fields. A comprehensive description of the Jovian rings-small moons system and ring-moon interactions within has been given by \citet{Burns2004}.

The ring-small moons systems of ice giants have recently been reviewed \citep{Showalter2020-mv}. Uranus is host to a total of 13 small inner moons, whose radial distances range from the edges of the main ring to nearly the orbit of Miranda, the innermost ``classical'' moon. From inside out, one finds first Cordelia and Ophelia, separated by the inner edge of the main rings. Then a group of 9 moons, as close to one another as a 3 percent fraction of their semi-major axis, forms the most densely packed set of moons in the Solar System. Finally, Puck and Mab orbit between this group and Miranda, surrounding a tenuous ring called the $\mu$ ring.

Taken together, the total mass of these small moons would form a dense ring, if fragmented. Their packed configuration is the seat of many resonances and instabilities, to the point that its lifetime is likely not larger than one million years. This made Showalter \citep{Showalter2020-mv} conjecture about the relationship between the lifetime and the age of this system, in an apparent paradox: ``The inner satellite system of Uranus presents us with a challenging puzzle. On the one hand, we have numerous indications that the system is unstable, with catastrophic collisions expected over time scales of approximately a million years or less. Cometary impacts are also capable of breaking apart small moons. On the other hand, we would expect the system to have formed, along with Uranus itself, over 4 billion years ago. How can we reconcile these two ideas? Perhaps the material that makes up these moons and rings is primordial, but its current configuration into this specific set of moons and rings is not.'' While the main point of this analysis still holds, it has been moderated by \citet{uk2022} who revisited the stability of the inner moons of Uranus using updated orbital elements and considering tidal dissipation. They found that the Belinda group can be stable on 10 8 yr timescales due to an orbital resonance between Belinda and Perdita and that, while tidal evolution cannot form the Belinda–Perdita resonance, convergent migration could contribute to the long-term instability of the Portia group. Additionally, recent modeling of the bombardment history of the giant planets \citep{Bottke2024} suggests that these inner moons might not have experienced impacts that would result in catastrophic disruption, at least for the largest of them. Finally, in the event a moon would experience significant disruption, the debris would likely re-accumulate rapidly into a new moon (or possibly several smaller moons), and/or be absorbed by nearby other moons, depending on the degree of radial dispersion of the debris. These analyses leave us with the idea that, while the material forming the small moons may be dating back to the giant collision that tilted Uranus' spin axis, their current configuration may be permanently evolving, redistributing material between objects and in radial distance over a broad range of timescales.

Neptune's six small moons offer a slightly different picture. While the four innermost ones orbit close to several of Neptune's narrow rings, the two outermost ones orbit outside of them. Being less packed than the small Uranian moons, these moons do not offer the same puzzle as Uranus, it has been suggested that they may be the result of another catastrophic event in the early solar system: according to the dominant theory, any primordial Neptune moon system may have been disrupted by the capture of Triton on a retrograde, highly eccentric orbit. Triton's orbit may have been progressively circularized afterwards, partly via tidal interactions with the planet, and partly by collisions with and accretion of the preexisting moons system Triton's capture contributed to disrupt, as suggested by \citet{Rufu_Canup_2017}. In this scenario, the tiny moon system we see today may have formed by re-accretion of the material of the putative primordial moons after completion of this circularization \citep{Agnor2006-ss}. It should be noted, however, that there is no definite proof of this scenario. From the fact that Neptune's small moons' sizes are comparable to those of Uranus, their formation via a scenario similar to the case of the small moons of Uranus and Saturn should not be totally discarded. 

Saturn's stands out in this family portrait by the diversity of its moons: excluding the thousands of very small moons embedded into the main rings and the 7 major regular moons, one can count 17 small regular moons. A large fraction of them occupy special orbit configurations; four on a Trojan orbit with larger moons, three of them are in co-rotation eccentricity resonance with Mimas \citep{hed10}, two moons shepherding Saturn's F-ring, two mutually co-orbital moons, and two more orbiting within gaps in Saturn's rings. Hyperion is locked in an orbital resonance with Titan. All other small regular moons orbit near the outer edge of the dense A Ring (as best seen in Figure \ref{fig:ring-moon}).
The diversity of these configurations points to a diversity of formation scenarios, which will be extensively discussed in Section 3. 

\subsection{Irregular moon systems}\label{irregular}
\begin{figure}[htbp]%
\centering
\includegraphics[width=0.9\textwidth]{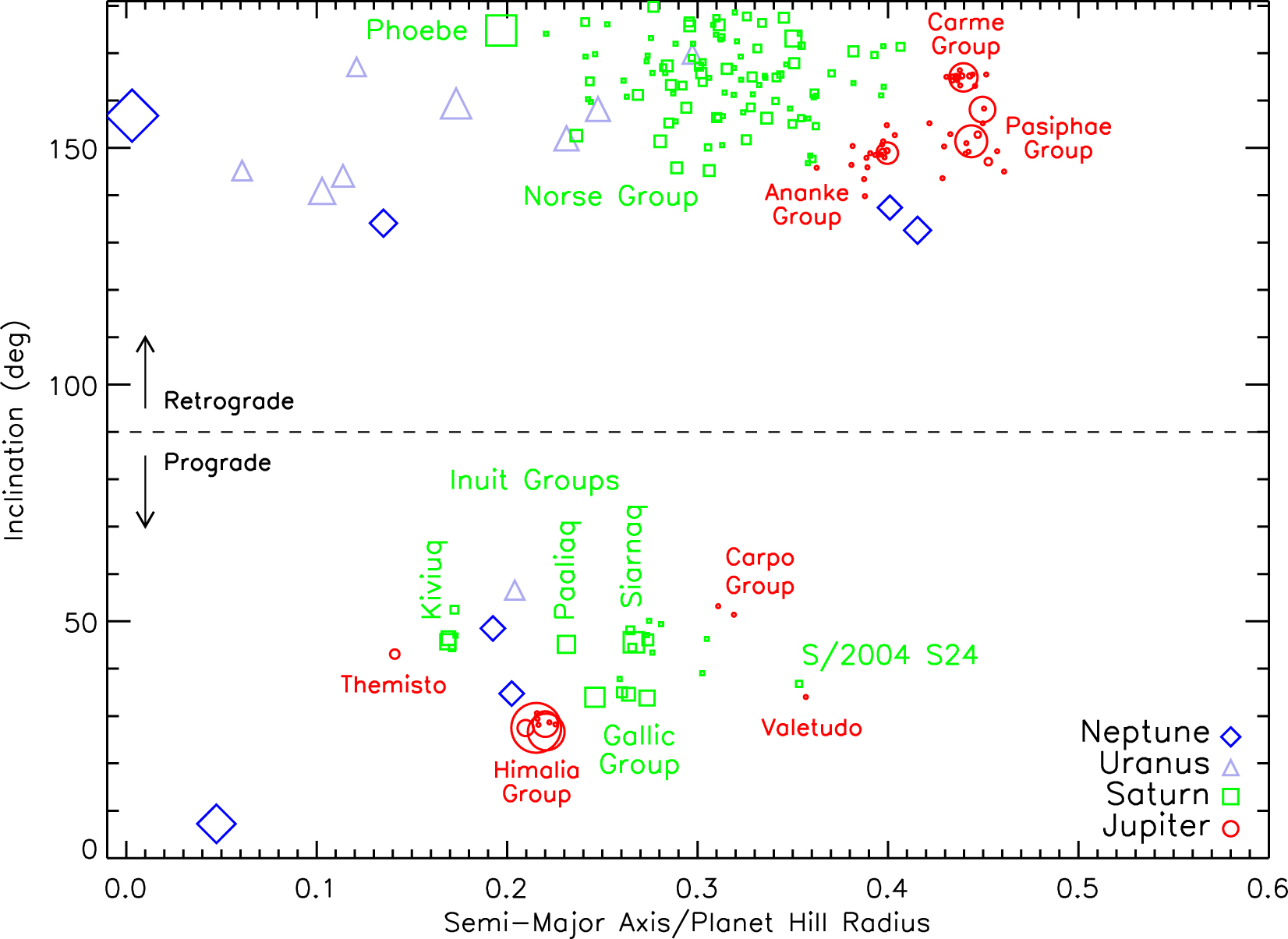}
\caption{Distribution of irregular satellites at Jupiter, Saturn, Uranus, and Neptune in a plane in which the horizontal axis is the semi major axis expressed in units of the planet's Hill radius, and the vertical axis is orbital inclination. Retrograde satellites appear in the upper part at inclinations above 90 degrees. In the top left corner one can see Phoebe, the only object that has been resolved and studied in detail by Cassini, and Triton, the only large moon of Neptune. The figure clearly shows the grouping of a large fraction of these objects in compact dynamical families (adapted from \citet{Sheppard2023-re}).} \label{fig:irregular}
\end{figure}

While most regular satellites occupy near-circular orbits deep inside the potential wells of their host planet, a population of small irregular satellites can be found around all four giants at larger distances, up to one half or two thirds of the Hill radius (see \citet{2008ssbn.book..411N} and \citet{Jewitt2005} for comprehensive reviews). Figure \ref{fig:irregular}, adapted from \citet{Sheppard2023-re}, shows the main characteristics of the distribution of orbital parameters of these populations, using a different color coding for each planet: preferred inclinations mainly between 30$^\circ$ and 60$^\circ$ (prograde) and between 130$^\circ$ and 170$^\circ$ (retrograde); rather large eccentricities spanning from 0.1 to 0.5, and strong tendency for a clustering in compact dynamical groups, such as the Ananke group for Jupiter or the Norse group for Saturn. The distribution of these irregular moons between prograde and retrograde orbits is also worth noticing. Figure \ref{fig:irregular} shows that a majority of the population is retrograde, but \citet{Bottke2010}, studying this distribution as a function of moon size, showed that retrograde moons are statistically smaller, while prograde orbits dominate among the largest of these objects, at and above 50 km diameter. They used a dynamical model of moon populations to show that this intriguing size distribution may result from the collisional re-shaping of an original population of objects similar to the Trojans captured during the chaotic episode of giant planets migration described by the Nice model. 

Saturn again occupies a special place in the inventory. First, the Cassini exploration of the Saturn system allowed an inventory and characterization of a fraction of the irregular moons family: using the time available between observations of the main mission targets, a systematic photometric study of irregular moons provided light curves over a broad range of phase angles for 25 of these previously poorly known objects \citep{Denk2019-ue}.   

Among them, Phoebe, the first irregular moon of Saturn, discovered as early as in 1899, was the object of a targeted flyby by the spacecraft that provided a 90 percent imaging coverage of its surface, revealing a dark (albedo 6 per cent), heavily cratered and near-spherical body; spectroscopic IR studies of its composition revealed the presence of water ice, carbon dioxide, phyllo-silicates, organics and possibly iron-bearing minerals at its surface \citep{Clark2005-cf}. Analysis of its density, indirectly related to its porosity, and of its implications on the history of its accretion, led to question the origin of this moon as a captured KBO \citep{Castillo-Rogez2019-oa} and to suggest that it could rather have formed from the same reservoir that gave birth to C-type asteroids.

Furthermore, thanks to spectacular progress made in recent years in telescope surveys of its Hill sphere in search for new moons, particularly at the CFH telescope on Mauna Kea \citep{Denk2019-ue}, Saturn is today the planet with the largest number of known moons, with a total of 146 including 122 irregular moons discovered as of 2023 (see \citet{Sheppard2023-re} and the ``green'' squares of Figure~\ref{fig:irregular}). These irregular moons generally have featureless visible and near infrared spectra dominated by water absorption bands. They are neutral or moderately red in color, pointing to some similarity with C-type, P-type or D-type asteroids, as well as with Jupiter Trojans. This similarity arguably points to a commonality in the capture mechanism for all of those populations, such as proposed by the giant planet instability scenario of the early Solar System.


\subsection{Regular moon systems}\label{regular}
\begin{figure}[htbp]%
\centering
\includegraphics[width=0.9\textwidth]{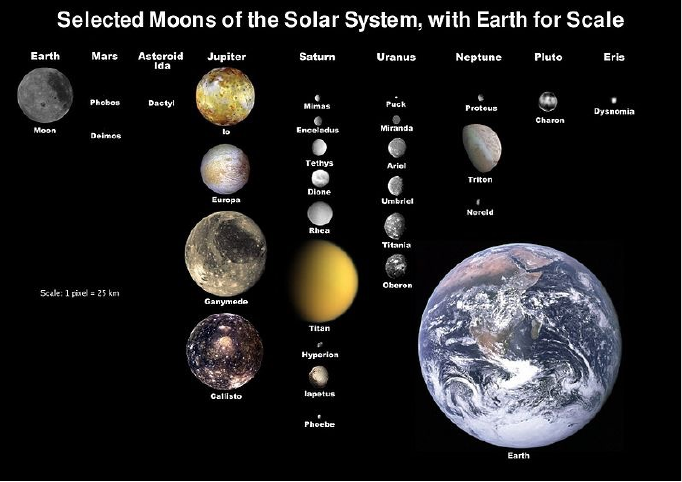}
\caption{Main moons in the solar system, with their sizes to scale.}\label{fig:regular}
\end{figure}

A notable characteristics of giant planets of our Solar System is the presence of systems of regular moons orbiting each of them. To analyze them first individually, Figure~\ref{fig:regular} shows the main giant planet moons placed in the broader context of the moons of all solar system planets, with Earth to scale. The figure reveals that each system displays deeply different characteristics. Jupiter stands out with its four large Galilean moons with comparable sizes not very different from the Earth's moon size. In contrast, Saturn has only one large moon, Titan, which gathers just by itself $96\%$ of the total cumulative mass of its 146+ moons. Then Uranus has no large moon but instead 5 medium-size moons with diameters similar to those of Saturn. Finally, while all other major moons occupy low-eccentricity prograde orbits close to the equatorial plane, Neptune's Triton, which represents just by itself $99.7\%$ of the mass of the Neptunian moon system ($2.1\cdot 10^{-4}$ Neptune mass), resides on a nearly circular but retrograde orbit with $156^\circ$ inclination best explained by a capture origin, hence its classification in the family of irregular moons (see previous paragraph).

As long as these moons are thought to have primarily formed in situ, their ubiquity has often been considered as indicative of their status of by-products of the formation of their parent planet \citep[e.g.,][]{canup2006common}. However, we will see in this article that in situ formation scenarios themselves cover different paths, and that capture is possible for some of them. Some additional observational characteristics of moons and their systems are particularly important to help distinguish between the different scenarios.

First, there are important characteristics to note in the overall architecture and physical parameters of each system of regular moons. As emphasized by \citet{canup2006common}, a common characteristic among Jupiter, Saturn and Uranus is the range of the total mass of regular moon systems, always between 1 and 3 times 10$^{-4}$ the mass of their parent planet. In contrast, the individual masses of moons span a broader range, from tiny objects to 10$^{-4}$ times their planet mass, the record being for Titan with $2.4\cdot 10^{-4}$ times the mass of Saturn.

\begin{figure}[htbp]%
\centering
\includegraphics[width=0.9\textwidth]{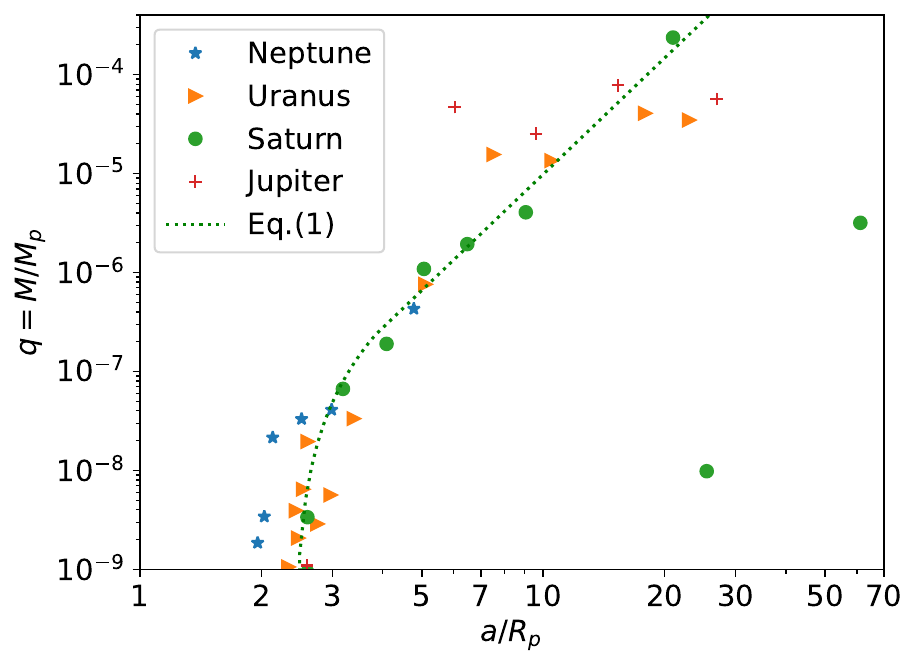}
\caption{Architecture of the regular moon systems of the four giant planets (Triton is not considered regular and thus not shown). Shown are the masses of the moons (normalized to that of their parent planet) against their semi-major axis (normalized to the radius of their parent planet). 
The dotted line corresponds to Eq.~\eqref{eq:q-Delta} (\citet{cri12}, see section \ref{massivedisk}), with a coefficient of proportionality equal to $2.3\cdot 10^{-7}$ and a Roche radius at 2.4 $R_p$ (140\,000 km for Saturn).} \label{fig:architecture}
\end{figure}

Another characteristic of the architecture of these regular moon systems is the variation of their mass with their distance to the planet, presented in Figure~\ref{fig:architecture} in units of planetary radius and mass. This diagram reveals a systematic increase of mass with radial distances at lower masses, which works particularly well for Saturn, starting from 0 at $\sim 2-3\,R_p$. This tendency reaches a plateau at large masses, in the mass range where both Titan and the four Galilean moons reside. The common trend observed in particular for the three outermost planets suggests a common process at play.

\begin{figure}[htbp]%
\centering
\includegraphics[width=0.9\textwidth]{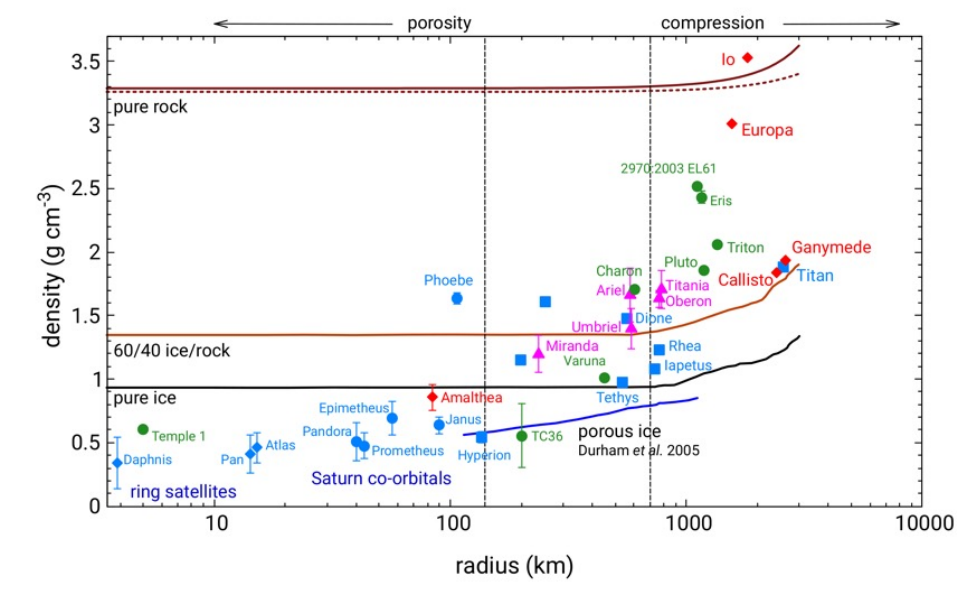}
\caption{Variation with radius of the density of outer solar system regular moons. Most of the density estimates used here are from spacecraft observations of radius and mass derived from gravity science analyses of flybys. Data for Ganymede and Callisto (orange diamonds), Uranus satellites (purple triangles), Triton (green circle) and Titan (blue square) are from \citet{yod95}. Amalthea data are from \citet{and05}. Data for mid-sized and small Saturn moons (radii $\sim10-1000$ km, blue squares) are from \citet{jac04,jac06,por05a,por05b,por07,tom07}. Pluto, Charon and KBO (green circles) data are, for Pluto/Charon from \citet{bui06}, Eris \citep{bro07}, Varuna \citep{jew02}, 2003EL61 \citep{rab06}, 1999TC36 \citep{sta06}. The curves labeled ``60/40 ice/rock'' (by mass) and “Pure Ice” are theoretical calculations of densities of a sphere with a rock-ice (`rock' density = 3662 kg m$^{-3}$) mixture and pure water ice composition, taking into account the presence of high density phases of water ice in the interior of a larger body \citep{lup79}. The curve labeled ``Porous Ice'' represents the density of an ice sphere with a porosity throughout equal to that allowed at the central pressure of the object, based on laboratory compression experiments with cold ice compaction \citep{dur05}. Variation of rock density with moon size for the ``pure rock'' case is expected to be very small; it was approximated using two EOS, one from \citet{mue93}, and one for \citet{poi00}, for a rock composition corresponding to a 3.2 to 3.3 g/cc density. Adapted from \citet{joh09}.}\label{fig:density}
\end{figure}

Finally, the density-radius relationship of the group of large moons (Figure~\ref{fig:density}) provides strong constraints on their bulk density and on their composition. For sizes below about 100km radius, they are less dense than water ice, suggesting a high degree of porosity as well as a large water content. For increasing sizes, their representative densities reach the density for water, then of rock-water mixes, suggesting not only a composition change but also a decrease of porosity with increasing mass and gravity. The large moons Io and Europa, among those with radii larger than 1000 km, stand out with densities near pure rock.

The concurrent variations of the density and of the mass of regular moons with distance to the planet provide important constraints on their formation (in situ or by capture) and on their evolutionary path, including the evolution of their volatile-to-heavy elements ratio from the time of their differentiation to contemporary times, which will be discussed in detail in Section 3 for the case of Saturn. 

\subsection{Specific features of Saturn's moons in the context of giant planets systems}\label{specific}

General reviews of the moons of the outer Solar System have been presented among others by \citet{Johnson2005} at the eve of the Cassini mission, and just after by \citet{Dougherty2018}, who reviewed the tremendous breakthrough in our knowledge of Saturn's moon system accomplished during the 13 years in orbit of this historic mission between 2004 and 2017. Saturn's rings and moons system clearly stands out by the large number of its objects and their broad diversity in mass, size, radial distance and composition. This system extends in radial distance from regions where moonlets accretion and disruption is still at play, inside the Roche radius for ice in the main rings, through intermediate distances dominated by gravitational dynamics, orbital evolution and resonances, to distant regions, a fraction of the Hill radius, where one finds the largest population of known irregular moons in the Solar System. Let us focus on regular moons, which offer the greatest challenge to our understanding of their formation.

First, inside Saturn's dense rings, one finds a large and diverse population of embedded moons, revealed by Cassini, such as the two small moons Pan and Daphnis which orbit inside two gaps of the main rings, respectively the Encke and Keeler gaps. Then, in-between rings and irregular satellites, 20 regular moons strike by their diversity in sizes, radial distances and dynamics. In order of increasing radial distances, one first finds five small moons: Atlas, Prometheus, Pandora, Janus and Epimetheus; the two last ones, located on very close orbits, are dynamically coupled to the point that they exchange their positions every fourth year. Then five mid-sized moons (Mimas, Enceladus, Tethys, Dione and Rhea) orbit inside of Titan, while two others (Hyperion and Iapetus) orbit outside Titan. Hyperion is locked in a 4:3 mean motion resonance with Titan. All these moons nicely align on the mass-distance average trend revealed by Figure~\ref{fig:architecture}, except for the two outer ones Hyperion and Iapetus which can be seen as the two green dots well below the main trend line in that figure. Finally, to be comprehensive, one finds an additional family or small moons orbiting in the region of mid-sized inner moons. The triad of Alkyonide moons (Methone, Anthe and Pallene) orbit between Mimas and Enceladus, and four additional small moons are Trojan companions of Tethys (Telesto and Calypso) and of Dione (Helen and Polydeuces). Finally, in-between Rhea and Hyperion stands Titan, Saturn's unique large moon and also the only moon in the Solar System with a dense atmosphere.

\begin{figure}[htbp]%
\centering
\includegraphics[width=0.9\textwidth]{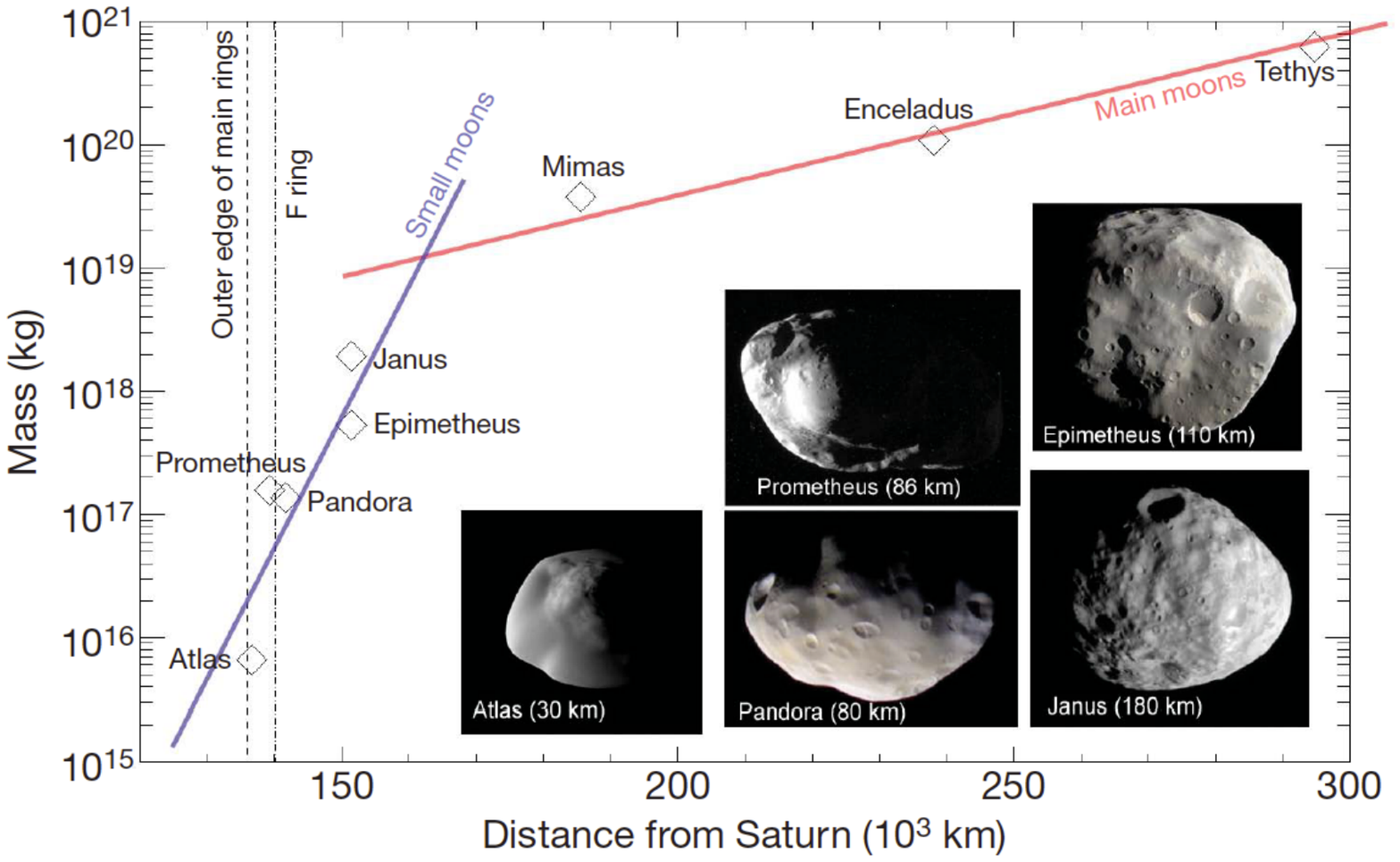}
\caption{Mass vs distance relationship for Saturn inner moons. One can note the marked change of slope between small moons (from Atlas to Janus, blue regression line) and mid-sized ones (from Mimas to Tethys and beyond, red regression line). From \citet{charnozetal2010}. Courtesy NASA-JPL-SSI.}\label{fig:Saturnmoons}
\end{figure}

Despite this intricate superposition of small and mid-sized regular moons, the same general trend already noticed in Figure~\ref{fig:architecture} emerges from their mass vs. radial distance relationship, as illustrated in Figure~\ref{fig:Saturnmoons}: a systematic increase of moons mass with planetocentric distance, with two different characteristic slopes for small inner moons (blue trend line) and for mid-sized ones (red trend line). 

Taken together, there are three main features that moon formation models have to reproduce at Saturn:

\begin{itemize}
    \item the systematic trend found on the mass-distance relationship of small inner moons and mid-sized moons;
    \item the presence of one single large Moon, Titan;
    \item and finally the presence of two mid-sized moons outside Titan's orbit. 
\end{itemize}

In the next section, we are going to explore the broad diversity of proposed formation mechanisms to try and assemble the puzzle of the Saturn moons system. 

\section{How did Saturn's moons form?}\label{saturnian}

\subsection{Introduction}\label{intro3}

\begin{figure}[htbp]%
\centering
\includegraphics[width=0.9\textwidth]{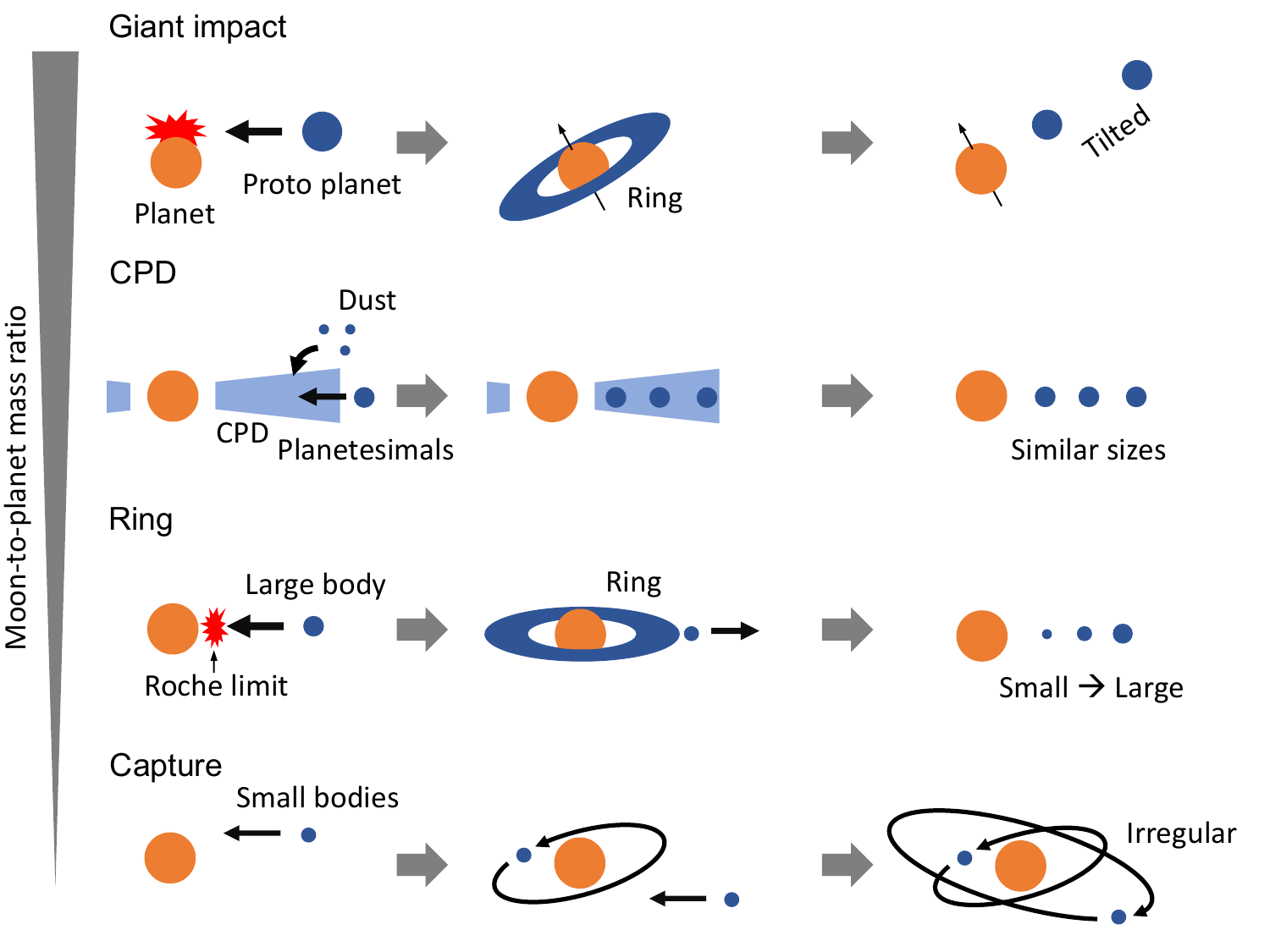}
\caption{An illustration of the four main mechanisms of moon formation showing their typical characteristics. The moons in the solar system were likely formed by one or combinations of these four elementary mechanisms, which are shown from top to bottom in the order of the typical moon-to-mass ratios of the systems that they can form.
}\label{fig:mechanisms}
\end{figure}

In order to best understand the specific case of Saturn and identify the most credible scenario(s) for the formation of its moons, we start by reviewing the diversity of moon formation scenarios currently considered for Earth, Mars and the four giant planet systems. The main four families of scenarios are illustrated in Figure~\ref{fig:mechanisms}, from top to bottom in the order of the typical moon-to-mass ratios of the systems that they can form.

As one will see, the age of moons, when it is known, can be a strong discriminant between the different scenarios, since it points directly to the period of their formation. Information about moons can be partly derived from their cratering record, with due consideration of the evolution of their surface after formation. In this respect the article by \citet{Bottke2024} which describes a comprehensive scenario for the bombardment history of the giant planet satellites, and references therein, provide important constraints for the ages of their different moon systems which are very complementary to the dynamical considerations we primarily develop in this section. In particular, the formation scenarios of mid-sized moons must take into account that their surface ages are on the order of 4 Billion Years old according to the cratering records, provided most craters are made by heliocentric projectiles.

\begin{itemize}
    \item formation by accretion in a debris disk formed by a giant impact or another type of catastrophic event: this is currently the favored scenario for the formation of Earth's Moon, recently reviewed by \citet{Canup2023}. It was also proposed for the two moons of Mars, Phobos and Deimos \citep{Rosenblatt2012}, and in a different way for Uranus by \citet{Morbidelli2012} to explain the formation of its system of mid-sized, prograde regular moons. These authors conjectured that the current Uranian moon system could result from the re-accretion of debris of a primordial moon system (similar to the Jovian one) disrupted by the giant impact(s) that likely tilted the spin axis of the planet. This scenario was later analysed in detail by \citet{Salmon2022}, who found that a set of specific conditions are needed, as well as some ad-hoc assumptions, in order for this scenario to work at Uranus. We will see in Section \ref{collisions} that this mechanism, i.e. re-accretion of a ``new generation'' of moons from a debris disk, has also been considered for Saturn's mid-sized moons.
    \item Formation from a ``primordial'' circumplanetary disk (CPD) is the favorite candidate for the Jovian system, dominated by its four large Galilean moons of similar sizes \citep{mos03a,canup2006common}. This generic CPD-related model may be applied to the primordial formation of prograde, regular moons occupying orbits near the equatorial plane of their host planet around a planet. It may be that it applies only to Jupiter in the Solar System, as the discussion of its applicability to Saturn will suggest in Section \ref{CPD}.
    \item Formation/migration from a massive ring: this scenario, initially proposed for Saturn owing to its large dissipation \citep{cha11}, before being generalised to the ice giants and the Earth \citep{cri12}, will be discussed in detail in Section \ref{massivedisk}.
    \item Capture of small bodies from Sun-centered orbits has been proposed by \citet{Pollack_1979} and \citet{Colombo_Franklin_1971} to explain the population of irregular moons of all four giant planets. For a capture to occur, a collision or a close encounter with a third body must be involved, such as the close encounters between giant planets proposed in the Nice model \citep{nes07}. Another possibility is to assume tidal disruption of a binary TNO by the capturing giant planet \citep{Agnor2006-ss}. For the Neptune system, current models propose that Triton's capture from the population of Trans-Neptunian Objects (TNO) and its subsequent circularization may have destroyed any primordial moon system. Thus the capture hypothesis for Triton concurrently explains the absence of other large or mid-sized moons around Neptune. Note however that capture of the small irregular moons of Jupiter, Saturn and Uranus does not seem to have perturbed their regular moons systems.
\end{itemize}

\begin{figure}[htbp]%
\centering
\includegraphics[width=0.9\textwidth]{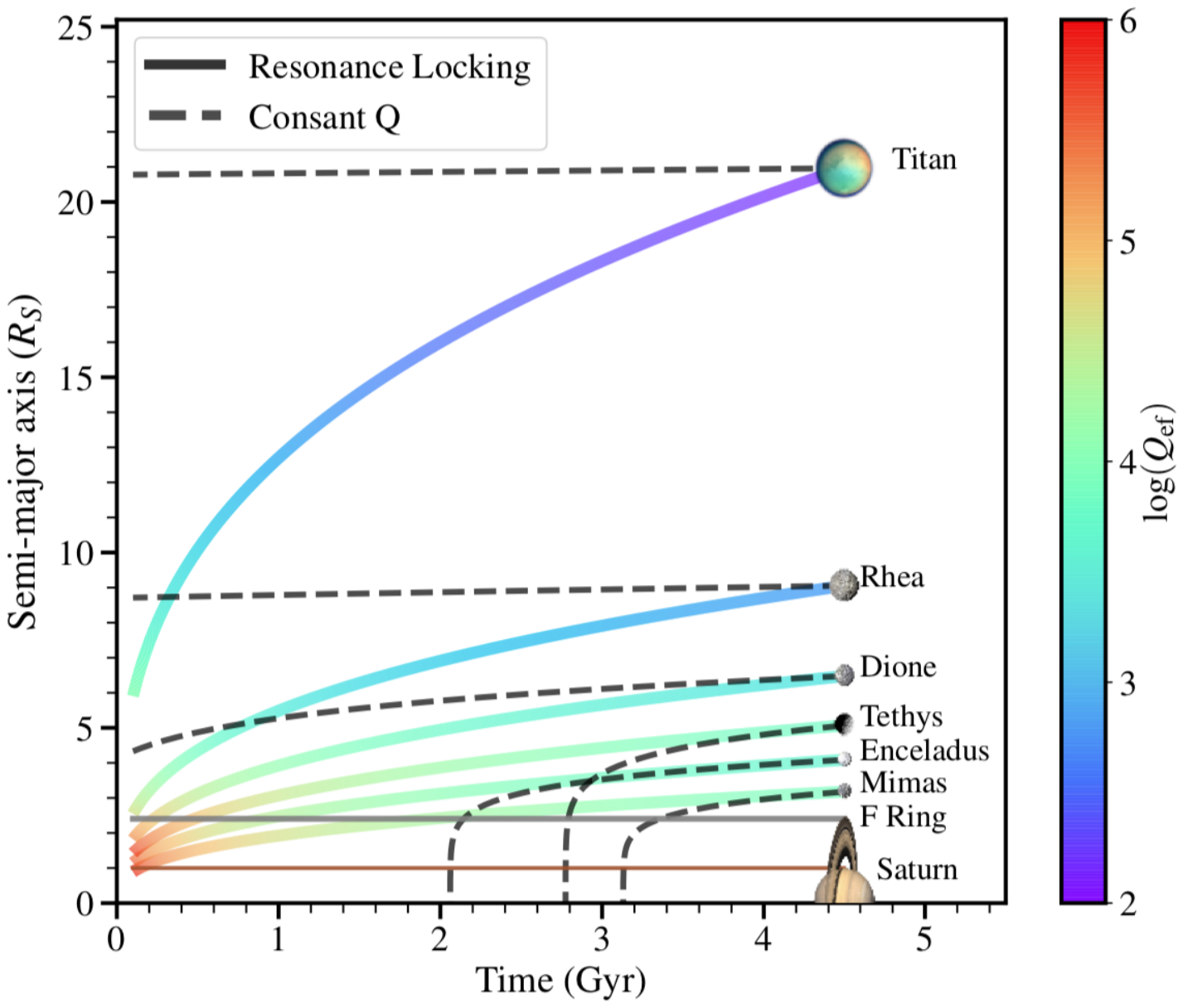}
\caption{This figure, originally from \citet{Fuller2016}, displays the expected radial migration trajectories over the age of the Solar System for two different types of tidal interactions driving migration. In the ``classical'' tidal interaction theory, the Q factor is basically constant and moon migration slows down quickly with increasing planetocentric distance (dashed curves); in contrast, tidal migration driven by propagting tidal waves and resonant locking (thick continuous curves) leads to similar migration time scales, on the order of the age of the Solar System, for all moons. See \citet{lai20} for a more detailed analysis.}\label{fig:migration}
\end{figure}

In the following sections, we are going to successively discuss how these different formation scenarios have been applied to Saturn's regular and irregular moons. These scenarios describe the resulting moon systems, but also introduce constraints on the locations where these moons have been assembled or delivered. In the context of a “highly dissipative Saturn” \citep{lai20}, these locations differ significantly from their current locations as a result of fast orbital migration. Determining the migration rate of each moon thus becomes a key element to retrieve the time and place of its formation. This migration rate theoretically depends on the Q factor associated to the tidal interaction mode responsible for orbital migration. In the ``classical'' tidal interaction theory, each moon interacts with the stationary tidal bulge it generates in the atmosphere and interior of its host planet. For a basically constant Q, the migration rate decreases with increasing radial distance, implying that distant moons like Titan experienced very limited migration, while moons orbiting closer to the planet experienced large ones. This model has recently been challenged by \citet{Fuller2016} who noticed that, if one considers all possible oscillations of the planet excited by tidal interaction with a moon, certain propagating tidal modes may produce a ``resonance locking'' between moons and planet: with this ``resonant locking'' mechanism, concurrent temporal evolution of the radial migration of the moon and of the characteristic resonance modes of the planet while it cools down may produce a migration regime which is independent of radial distance and equally well applies to distant moons like Titan and to inner moons. The corresponding migration timescales are on the order of the age of the solar system for all moons, as shown in Figure \ref{fig:migration}: starting from the current location of moons and moving back in time, the ``classical'' tidal theory (dashed curves) predicts that Titan must have formed close to its current location whereas inner moons may have formed significantly later than the Solar System. In contrast, assuming resonant locking in tidal migration (thick continuous curves), all moons migrate at comparable rates over the age of the solar system. Recent measurements of Titan's orbit expansion rate by two independent methods \citep{lai20} tend to validate the predictions of the resonant locking model. Even so, further measurements are needed to ultimately discriminate between the different types of tidal interaction and orbital migration illustrated in Figure \ref{fig:migration} and to draw consequences on the validity of the different formation scenarios.

The reader should note that, while Figure \ref{fig:migration} captures the main point of resonance locking, (1) it simplifies it somewhat, as for instance the Mimas-Tethys and Enceladus-Dione pairs must evolve convergently to be in their present resonances, and (2) the actual migration rates traced by the solid curves of \ref{fig:migration} are functions of the model of Saturn interior used, and may be significantly different if one chooses another interior model. A more detailed description of the latest orbital migration theories can be found in Chapter 4 of this Topical Series (Cuk et al., 2024).

\subsection{Formation of the regular moons of Saturn.}\label{saturnian-regular}
In this section we successively review the different scenarios that may apply to the formation of Saturn’s regular moons and discuss their respective merits. 

\subsubsection{Late formation of mid-sized moons after disruption and re-accretion of a previous moon system}\label{collisions}

\begin{figure}[htbp]%
\centering
\includegraphics[width=0.9\textwidth]{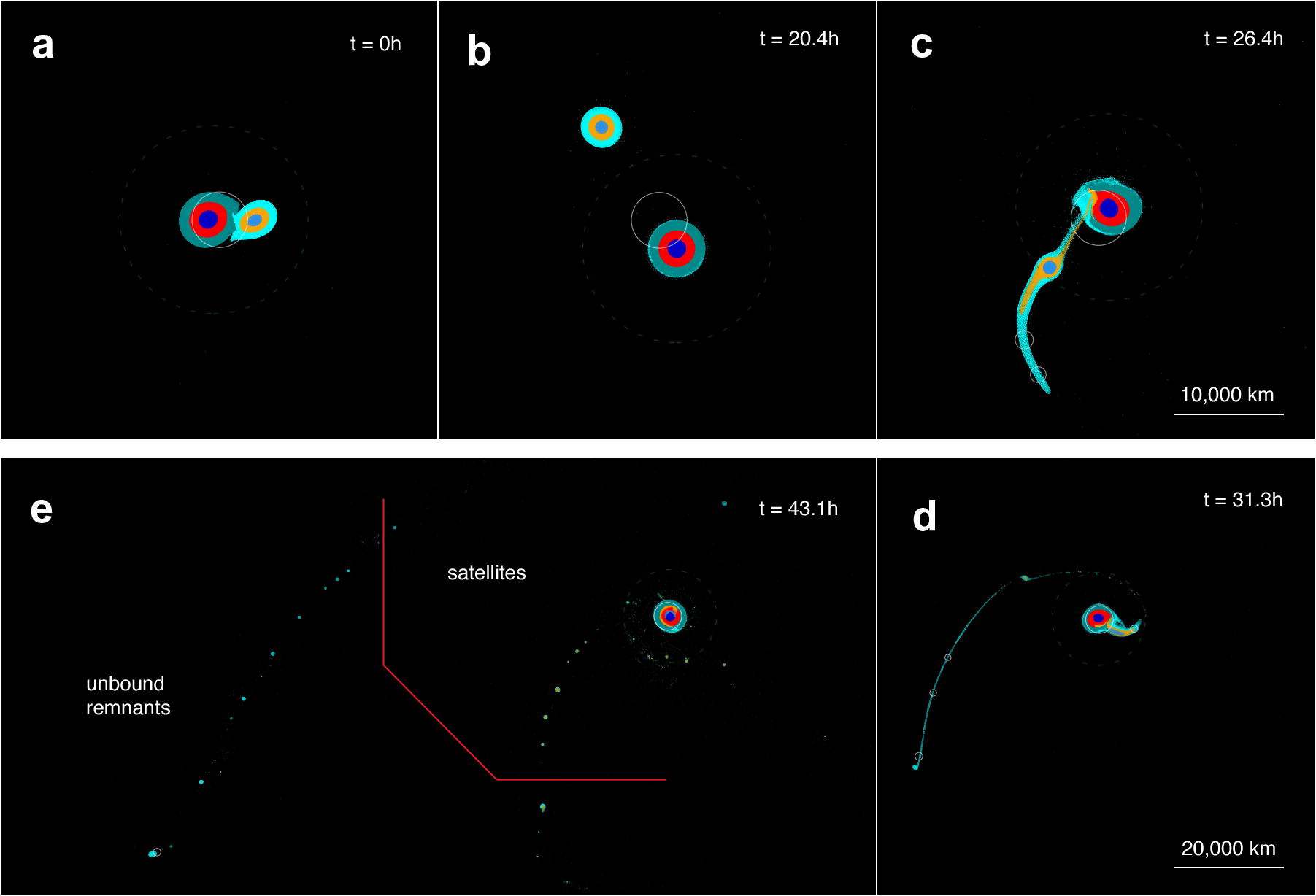}
\caption{Example of one of the giant collisions simulated by \citet{Asphaug2013} to simultaneously reproduce Titan (from merging of several smaller-sized bodies in successive collisions) and the mid-sized moons (from clumps of one or several of these collisions). Here five consecutive snapshots of a 1:3 mass ratio collision between a satellite of a mass comparable to Titan and a smaller impactor are shown clockwise from a to e. While a series of clumps escape the system, a few of them (right of the red broken line in panel e) remain in Saturn-centered orbit and form MSM's. In this scenario, Titan's formation bears some similarity with the oligarchic growth of terrestrial planets in the inner Solar System. See \citet{Asphaug2013} for more details.}\label{fig:fig7}
\end{figure}

This first family of scenarios proposes that Saturn's moons formed from the collisional disruption and re-accretion of a previous moon system. A first study along this line \citep{Asphaug2013} started from the initial assumption that Saturn's CPD, just like Jupiter's, gave birth to a moon system similar to the Galilean moons that was later disrupted during an episode of global instability. Following a detailed critical analysis of the physics of collisions between moons and the clumps of ice-rich material produced in these collisions, this study performed a full simulation of the moon formation sequence using a 3-D Smooth Particle Hydrodynamics (SPH) code, together with an equation of state describing the colliding differentiated bodies. In this simulation, Titan is formed at the same time as the unique large-size end-product of a series of mergers, in a way similar to the oligarchic growth of planetesimals that likely led to the formation of inner planets. Figure \ref{fig:fig7} shows a time series of the events occurring in one of these collisions. A merit of this scenario is that it simultaneously reproduces Titan and the mid-sized moons as the products of the same series of collision-disruption-merger events. However, the assumptions made on the properties of collisions and on the rheologic behavior of colliding moons and fragments are not easy to validate by current observations, and the model does not explain the mass-distance relationship of small moons and the circularization of the orbits of new-born moons after their re-accretion.

Another scenario of moon re-accretion from an earlier system was proposed by \citet{uk2016}, starting from an analysis of the role of orbital resonances in the dynamical evolution of mid-sized moons. Using two different numerical integrators, the authors studied two resonance crossings that should have happened in the past based on the current orbits: the 3:2 MMR of Tethys and Dione, and the 5:3 MMR of Dione and Rhea which should have happened later. Their simulations suggested that the Tethys-Dione 3:2 resonance crossing always produces excessive final inclinations for these moons which are incompatible with their present values. On the other hand, simulations of the Dione-Rhea 5:3 resonance reasonably reproduce current eccentricities.

If only the second resonance crossing happened, but not the first one, Dione and Rhea must have formed between the two potential resonance crossings. Taking into account the fast tidal migration observed by \citet{Lainey2012}, this implies that formation of this moons was recent, maybe as young as 100 Myr. Building on this ``young moons'' hypothesis, \citet{uk2016} proposed that Saturn's mid-sized moons formed from a disk of solids that briefly existed late in the history of the Saturn system from the catastrophic disruption of a previous moon system. In this scenario, this disk extending from inside to outside of the Roche limit could have been the progenitor of the current rings. Like the previous one, this model does not explain the circularization of small moons orbits and the mass-distance relation shown in Figure~\ref{fig:architecture}.

As noted by the authors, further validation of this scenario and of other similar ``young moons'' scenarios rests on three main types of evidence: 1) a better estimate of the tidal heating rate of Enceladus; 2) a possible post-Cassini reconsideration/validation of the fast tidal migration rates of Saturn's moons observed by \citet{Lainey2012},  in particular because in a typical resonance locking mechanism the period ratios do not evolve, preventing resonance crossing (see figure~\ref{fig:migration}), and 3) a careful analysis of the cratering records of icy moons, a critical element to date the age of their surfaces that requires taking into account the distribution between impactors in planetary and in sun-centered orbits. 

\subsubsection{Formation from a CPD}\label{CPD}
Regular moons of giant planets, particularly those of considerable size, are thought to originate within circumplanetary disks (CPDs), akin to how planets form within protoplanetary disks (PPDs). In our solar system, this phenomenon is evident with the Galilean moons of Jupiter, and possibly with Titan and Iapetus around Saturn. CPDs emerge as byproducts of the gas accretion process of planets, as the accreting gas carries angular momentum, allowing planets more massive than Saturn to carve gaps within the PPD \citep[e.g.,][]{cri06}. It is improbable that Uranus and Neptune harbored their own CPDs, as they likely remained engulfed within the PPD \citep{cri06}. In this section, we present an overview of proposed scenarios regarding the formation of large moons within CPDs. The dimensions of CPDs typically span approximately 0.3 Hill radius of their host planets (approximately 0.1 AU for both Jupiter and Saturn) \citep[e.g.,][]{lub99}. Moons are theorized to originate within the high-density regions situated close to the centers of these disks, typically at radial distances smaller than 0.03 to 0.1 Hill radius, as described in the majority of formation models.

Two classical types of formation scenarios have been delineated in the literature: the minimum mass model and the gas-starved disk model. In the first model, CPDs are segregated from their parent PPDs. The solid mass utilized for moon formation within these disks approximates their current total mass in the Jupiter, Saturn, and Uranus systems, denoted as $M_{\rm s}\sim10^{-4}~M_{\rm p}$. Within the Saturnian system, the mass contrast between Titan and Iapetus finds explanation in the distribution of solid masses; the majority of mass is concentrated within the centrifugal radius situated around Titan's current orbit, with only a marginal quantity of solids near Iapetus's current orbital position contributing to its formation \citep{mos03a,mos03b}.

The other classical scenario is the gas-starved disk model \citep{can02}. In this framework, gas accretion onto the CPD persists during moon formation, albeit at a rate modest enough to maintain the disk temperature below the sublimation point of ice across most regions of the CPDs, thereby facilitating the formation of icy moons. Solid material is consistently delivered to CPDs from their parent PPDs, with varying accretion rates assumed in hydrodynamic (HD) and magneto-hydrodynamic (MHD) simulations \citep[e.g.,][]{gre13,kan15b,mul18}.

Recent numerical simulations and observations of planets embedded in gaps in PPDs have revealed that modest gas accretion persists even after embedded planets have carved gas gaps along their orbits \citep{haf19,szu22}. This observation lends stronger support to the gas-starved disk model over the minimum mass model. Consequently, in the following we delve into recent models grounded in the gas-starved disk concept. A range of concepts rooted in this framework have been investigated to replicate the typical (total) satellite-planet mass ratio, $M_{\rm s}/M_{\rm p}\sim10^{-4}$, along with other characteristics of large moons.

\begin{figure}[htbp]%
\centering
\includegraphics[width=0.8\textwidth]{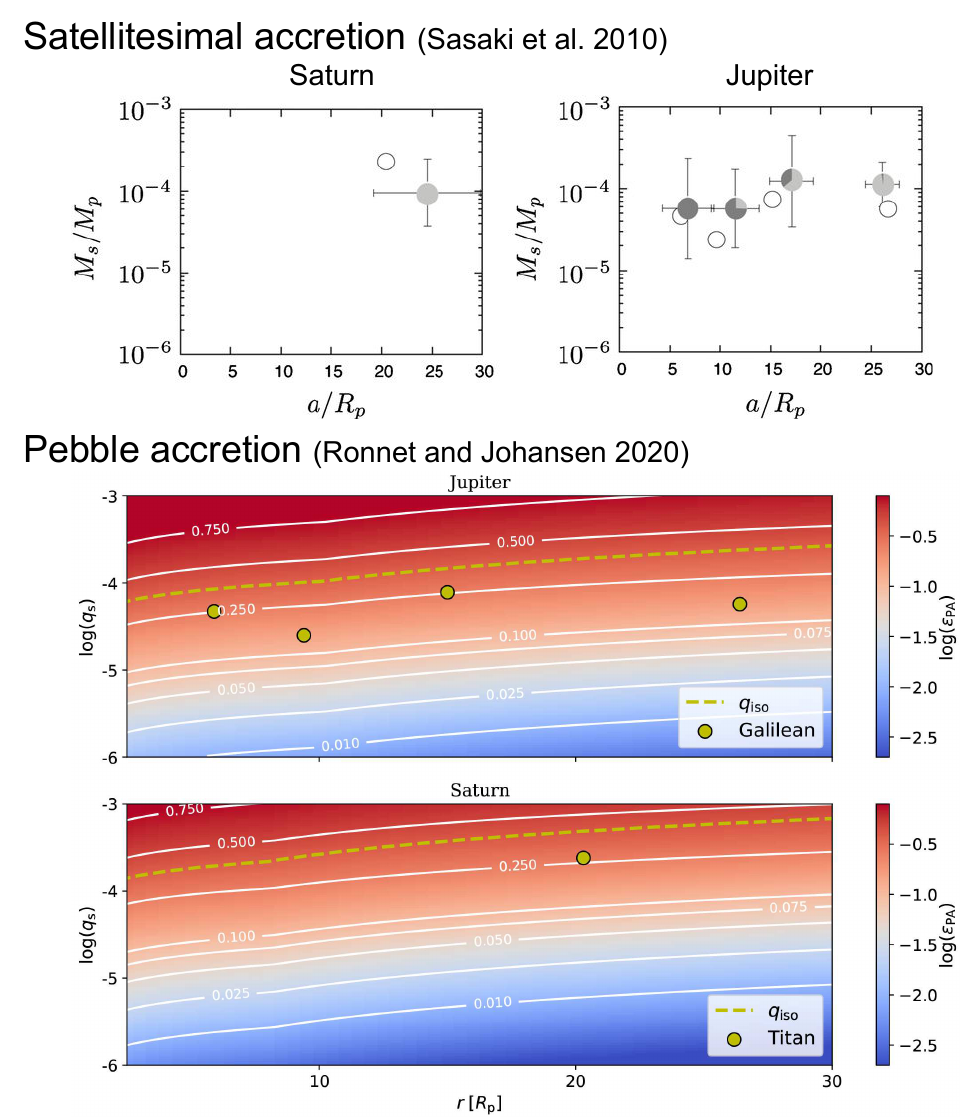}
\caption{Formation of the moons around Saturn and Jupiter in the classical satellitesimal accretion and recent pebble accretion scenarios. The upper panels show that the mass and orbits of both Saturnian and Jovian moons are reproduced by population synthesis simulations in \citet{sas10}. The thickness of gray represents the rock-to-ice mass ratios, which are also reproduced well. However, the plenty of satellitesimals assumed to exist in the CPDs in the simulations could be problematic, as we explain in the text. The lower panels show that the current mass of moons (yellow circles) is lower than the pebble isolation mass (i.e. the satellite mass where pebble accretion stops by curving gaps around the satellites; yellow dashed curves), meaning that the moons can be formed by pebble accretion if their orbits are fixed. The color maps show the pebble accretion efficiency (the ratio of the mass of accreting pebbles to the mass of the satellites growing out of the drifting pebbles). See \citet{sas10,ron20} for more details.}
\label{fig:CPD_formation}
\end{figure}

In the original concept of the gas-starved disk model, multiple generations of large moons form successively within the CPDs. N-body simulations suggest that the satellite-planet mass ratio, $M_{\rm s}/M_{\rm p}\sim10^{-4}$, can be broadly replicated through a balance between their loss via Type I migration toward the planets and their formation through the collisional growth of satellitesimals \citep{canup2006common}, which are kilometer-sized building blocks of moons \citep{shi17}. Some subsequent studies proposed that the orbital positions of Io, Europa, and Ganymede, which are locked in the Laplace resonance, could be reproduced if the strong magnetic field of young Jupiter generated an inner cavity within the CPD, thus halting moon migration (refer to the upper panels of Fig. \ref{fig:CPD_formation}) \citep{sas10,ogi12}. Conversely, young Saturn, being smaller than young Jupiter, likely lacked the capacity to generate a sufficiently robust magnetic field to carve out a wide inner cavity at the CPD's center. This scenario would have led to the survival of only one large moon and perhaps the existence of massive ice rings (see Section \ref{tentative}).

However, a critical challenge with this scenario lies in the formation of satellitesimals. As dust particles supplied to the CPDs increase in size through mutual collisions and adhesion, they tend to drift towards the planets as centimeter to meter-sized pebbles before reaching the size of satellitesimals, akin to the movement of dust particles in PPDs towards stars before they can evolve into planetesimals \citep{shi17,zhu18}. Three primary solutions have been proposed to address this issue: 1) outward gas flows interacting with dust and pebbles in specific regions facilitate the formation of satellitesimals via the streaming instability, 2) planetesimals formed within PPDs are captured by CPDs and assume the role of satellitesimals, and 3) dust particles efficiently grow within laminar magnetic wind-driven CPDs.

The operational mechanisms and characteristics of outward gas flows within CPDs remain a subject of contention. According to HD and non-ideal MHD simulations, such flows may manifest in the midplane of CPDs, particularly in the outer regions of the satellite formation zone, contingent upon specific viscosity and opacity conditions \citep{tan12,gre13,szu14,szu17b,sch20}. Outflows could aggregate particles at the orbital location of approximately 0.1 Hill radius or beyond, fostering a consistent generation of satellitesimals via the streaming and/or gravitational instability, potentially resulting in successive generations of moons \citep{dra18b,cil18,bat20}.

The capture of planetesimals has also been explored in various studies \citep[e.g.,][]{rus06,sue17}. Gas pressure bumps emerging at the outer perimeters of gaps carved by young Jupiter/Saturn within PPDs could serve as efficient sites for planetesimal formation \citep[e.g.,][]{lyr09,shi20}. Conversely, since planetesimals are also pushed away from the vicinity of planets due to gravitational interactions as gap structures form within the gas, capturing planetesimals by CPDs necessitates high eccentricity for them to approach planets closely enough \citep{fuj13}. Furthermore, the migration of an outer planet can perturb planetesimals around the inner planet, potentially explaining why Jupiter, affected by Saturn's migration, boasts four large satellites while Saturn has only one \citep{ron18,eri22}.

Recent non-ideal MHD simulations have also demonstrated that CPDs can function as laminar magnetic wind-driven disks. Within such laminar disks, dust particles can settle onto the midplane more efficiently compared to the turbulent CPDs assumed in prior studies, allowing for more efficient growth through mutual collisions. Consequently, satellitesimals can form before dust descends towards the central planet \citep{shi23}.

Pebble accretion presents an alternative mechanism for moon formation. Initially proposed in the context of planet formation within PPDs, pebble accretion utilizes drifting dust directly as the building material for planets \citep{orm10,lam12}. Here, \emph{pebbles} are conceptualized as objects (dust aggregates) large enough to decouple from the gas (unlike dust grains) but small enough to be significantly affected by gas friction (unlike planetesimals)\footnote{Given that this definition corresponds to a Stokes number $S_{t}$ (friction stopping time multiplied by orbital frequency) on the order of \emph{one} regardless of size, \emph{s$_{t}$one} may be a more suitable term than \emph{pebble} \citep{cri23}.}. Pebbles typically range from centimeters to meters in size, depending on the density of the surrounding gas. Planetary embryos efficiently accrete pebbles drifting towards the central stars due to gravitational focusing effects and aerodynamic drag, which influences pebbles more strongly than planetesimals. Consequently, the feeding zone for pebbles is considerably larger than that for planetesimals. This mechanism operates analogously for the formation of large moons if their seeds are supplied to CPDs and their migration is halted by the inner cavity (refer to the lower panels of Fig. \ref{fig:CPD_formation}) \citep{shi19,ron20}. Captured planetesimals ablated by CPD gas can form pebbles through recondensation, and large planetesimals can also serve as seeds for moon formation \citep{ron20}.

Two potential challenges arise when applying pebble accretion to the case of Saturn. Firstly, the ablated material from which Saturnian moons accreted might lack ice, which contradicts their observed low densities. Nonetheless, if this material was abundant in hydrated minerals, the dehydration of particles and outward diffusion of released water vapor could have facilitated the recondensation of substantial amounts of ice within the formation region of the moons \citep{mou23}.

The other issue pertains to the necessity of inner cavities to impede rapid inward migration \citep{ron20}. Given that this condition likely wasn't met at Saturn, the formation of Titan through pebble accretion becomes challenging, or at best, it necessitates that this moon formed in the very inner region of the CPD before migrating outward. This scenario aligns with the recent observation of the current rapid outward orbital evolution of Titan: \citet{lai20} demonstrated that Saturn's moons migrate outward due to Saturn's tides at such a swift rate that it would take less than the age of the Solar System to move from the Roche radius to their current positions. This observation opens the possibility for a more intricate and dynamic history than previous formation scenarios, as outlined in Section \ref{tentative}.

\subsubsection{Formation/migration from a massive disk}\label{massivedisk}

We call \emph{tidal disk} a disk of condensate debris which is dynamically cold, but prevented from coalescence into self-gravitating objects by the tidal forces from the central planet. In other words, a tidal disk is a debris disk inside the Roche radius. Saturn's main rings are the best example of such a tidal disk, but the moon-forming disk of molten lava which probably surrounded the Earth is an other case.

Interactions among the disk material yield a radial spreading of the tidal disk. When a dynamically cold disk of solids (like Saturn's rings) spreads, self-gravity dominates over tides beyond the Roche limit of the planet, and solids aggregate into elongated objects. As they orbit around the planet, these objects can keep accreting material from the rings, since the Roche radius depends only on the density of the considered material, not on the size of the moonlet.

\begin{figure}%
\centering
\includegraphics[width=0.8\textwidth]{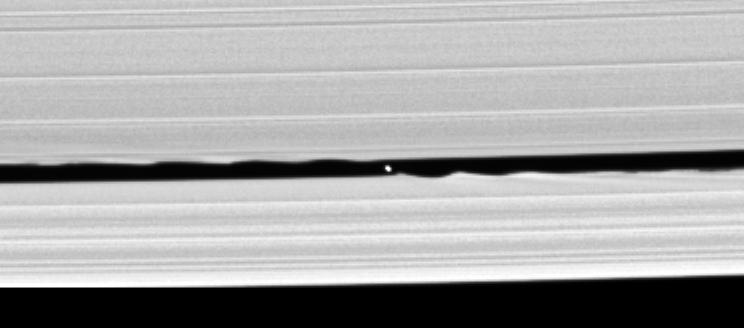}
\caption{Famous Cassini picture of Daphnis in the Keeler gap at the outer edge of the A-ring. The gravitational effect of the moonlet on the rings leads to the waves which can be seen trailing behind Daphnis outside its orbit, and leading ahead of Daphnis inside its orbit.}
\label{fig:Daphnis}
\end{figure}

However, moonlets also migrate outwards. Indeed, their orbital velocity is smaller than that of ring particles, so that the gravitational perturbation caused by a moonlet on the orbits of ring particles decreases the angular momentum of the latter, and increases that of the former. Ring particles lose angular momentum by gaining eccentricity, but are quickly circularised by collisions with other ring particles. Such a phenomenon can be observed in the vicinity of the ring moons Pan and Daphnis, and explains how these two satellites maintain their gaps open (see figure~\ref{fig:Daphnis}). Accumulating interactions with many ring particles, the moonlet gains angular momentum, and therefore is transferred onto a circular orbit of larger radius. The migration rate is proportional to the masses of the moonlet and of the rings, and to the distance to the edge of the rings to the power $-3$ \citep[see for instance][]{LinPapaloizou86}.

Depending on the outcome of the competition between the disk spreading rate and the moon migration rate, one could have a single moon absorbing all the mass from the tidal disk as it migrates away (which would be the Earth-Moon case) or a series of moons regularly popped out of the rings and migrating away at constant mass; in this case, since their migration rate decreases with distance to the rings, the separation between moons decreases and they 
suffer collisions and merge, leading to a mass increase with orbital distance.

This process has been modeled numerically by \citet{charnozetal2010} around Saturn's present rings and nicely reproduces the handful of small moons Janus, Epimetheus, Prometheus, Pandora and Atlas, which share the same spectrum as the rings and a low bulk density, strongly suggesting that they are made of the same material. Besides, the migration rate due to their interactions with the rings is such that it is impossible for them to have been on their orbit for billions of years. A numerical application shows that Janus should have been inside the Roche radius just a few hundred million years ago. There is now a consensus that these small moons are born recently, from the spreading of the present rings beyond the Roche radius.

This model has then been generalized to all the mid-sized moons of Saturn, using two key recent results. First, ring formation and evolution models \citep{can10,Salmonetal2010} suggest that Saturn's rings could be primordial but have been much more massive in the past. In this case, they could have lost a mass equivalent to most of the mid-sized moons over the age of the Solar System by spreading beyond the Roche radius. Second, the strong dissipation measured inside Saturn \citep{lai20} allows for a migration of the mid-sized moons to their present orbit from the Roche radius over the age of the Solar System \citep[see Appendix of][]{cha11}.
This is a key element in the discussion since migration due to the interaction with the rings vanishes past the location of the 2:1 mean motion resonance (MMR) with the Roche radius, that is $222\,000$~km from Saturn, between the orbits of Mimas and Enceladus. Hence, with the previously estimated quality factor for tides inside Saturn which makes migration over the age of the solar system negligible, it would be impossible to form the mid-sized moons by this process. Conversely, with the strong dissipation measured about 15 years ago, Saturn's tide can take over the interaction with the rings and keep pushing the satellites outwards. 
Starting from massive rings, numerical simulations reproduce quite nicely the distribution of all mid-sized moons up to Rhea \citep{cha11}. Besides, this model offers an explanation for the presence and variety of cores among these 5 moons, whose silicate to ice ratio varies from $7\%$ to $70\%$. In this model, silicates originally present in the rings would have aggregated into a few large objects within the rings (because the Roche radius for silicates is only $\sim 90\,000$~km). This handful of proto-cores of random masses would then have been covered with ice during the process of ring spreading, providing the seeds for Mimas, Enceladus, Tethys, Dione, Rhea. In this model, the 5 aforementioned moons were born in sequence, the further and more massive the earlier (thus the older the object is now), but they grew through multiple collisions after having left the rings.

Using an analytical description of this model, \citet{cri12} demonstrated that the spreading of a tidal disk beyond the Roche radius, in the case when it does not yield one single large moon, leads to the formation of a series of moons with a precise mass - distance relation (where the mass $q$ is expressed in units of the central planet mass, and $\Delta$ is the distance to the Roche radius, normalised by the latter):
\begin{equation}
q \propto \left\{
\begin{array}{ll}
(\Delta/\Delta_{2:1})^{9/5}                        & \mathrm{\ for\ } \Delta<\Delta_{2:1}\\
\left(\frac{\Delta+1}{\Delta_{2:1}+1}\right)^{3.9} & \mathrm{\ for\ }\Delta>\Delta_{2:1}
\end{array}\right.
\label{eq:q-Delta}
\end{equation}
where $\Delta_{2:1}\approx 0.58$ is the location of the 2:1 MMR with the Roche radius.
This analytical relation
nicely fits the one observed in the Saturn system, from Pandora to Rhea and even Titan; the observed masses are within a factor $60\%$ of the analytic relation across 4 orders of magnitude in distance (see figure~\ref{fig:architecture}).

It is interesting to note that each of these three results from around 2010 --\,(i) initial massive rings which spread and lose mass over the age of the solar system, (ii) formation of satellites at the Roche radius followed by migration to their present position and (iii) strong tidal dissipation inside Saturn\,-- are, alone, inconsistent with the present system. However, taken together, they draw a new consistent picture for the origin of the Saturnian ring and satellite system. But each relies on the other two. For instance, the strong tidal dissipation is not possible if the satellites have been here since Saturn formed. But reciprocally, the satellites can not have been formed in sequence from the spreading of the rings if they don't migrate fast.
And it is absurd to consider initially massive rings if their mass can not have been evacuated, or to imagine that the moons come from the rings if the latter can not efficiently spread and send mass beyond the Roche radius.

In this frame, the newest results on Saturn's tides, suggesting that instead of the constant $Q$ paradigm, the resonant lock mechanism may prevail \citep{lai20}, can modify the picture. Indeed, Eq.~\eqref{eq:q-Delta} relies on the constant $Q$ for $\Delta > \Delta_{2:1}$. But with the resonant lock, satellites migrate faster as they go further and do not merge anymore. The mass - distance relation would then be the consequence of a relation between the mass at $\Delta_{2:1}$ and the age, which remains to be established. On the other hand, the observed fast migration of Titan offers the possibility to include Titan in the process \citep{lai20}, while it was impossible to bring it to its position in 5 Gyrs with the constant $Q$ paradigm. So some work remains to be done to update this satellite formation model.

Even more interestingly, the relation \eqref{eq:q-Delta} matches the mass - distance distribution of the regular moons of Uranus and Neptune, suggesting that ice giants also had massive rings in the past \citep{cri12}, which should then have been later completely eroded by meteoritic bombardment. In that sense, the Saturnian system would be an illustration of a more general process: see Crida et al. (2024, this collection), for a more detailed description of this mechanism.
As one can see from Figure~\ref{fig:architecture}, this same relation doesn't work for Jupiter and the Galilean moons. But the case of Mars has also been shown to be consistent with this model, although indirectly: in one of their formation scenarios, Phobos and Deimos would be formed from debris accumulated in the MMRs with the most massive satellite coming out of the rings. All the satellites directly formed from the rings would then have already fallen back onto Mars due to tides \citep{Rosenblattetal2016}. However alternative scenarios have been proposed, such as for instance the direct formation of Deimos from a disk of giant impact debris that extended beyond the synchronous orbit \citep[e.g.,][]{can18}. 

Finally, let us note that the model described here does not address the origin of the tidal disk. In the case of the Earth, Mars, and maybe of Uranus, a giant impact is the favored hypothesis. In the case of Saturn, the ring formation model which best fits the scenario described here is based on the tidal stripping of the icy mantle of a Titan-mass satellite which migrated inside the Roche radius inside Saturn's CPD \citep{can10}. Hence, this satellite formation model actually relies on a first generation of satellites formed in the CPD as described in the previous section. We will propose a possible formation scenario of the mid-sized moons and Titan based on this idea in Section \ref{tentative}. The reader is referred to the review of the origin of the rings presented in Crida et al. (2024, this topical collection) for more discussion on this topic.

\subsection{Origin of irregular moons}\label{trregular}
Most of the irregular moons orbits have large inclinations to their planet's equator and a majority of them are retrograde (see Fig. \ref{fig:irregular}),  suggesting that they originate from Sun-orbiting small bodies that have been captured by Saturn during their crossing of its Hill sphere (see also Fig. \ref{fig:mechanisms}). Indeed, irregular moons may have been captured from previously heliocentric orbits by gas drag \citep{Pollack_1979}, collisions during the Solar Nebula period \citep{Colombo_Franklin_1971}, or three-body interactions during the final stages of the large-scale dynamical instability described in the Nice model \citep{2007DPS....39.3211N}. These different scenarios involve a diversity of capture mechanisms.

(1) “pull-down” capture during the short phase of rapid growth of a body like Jupiter due to the run-away accretion of its gas envelope \citep{Heppenheimer_Porco_1977}: the resulting fast expansion of its Hill sphere may result in the entry and subsequent capture of small bodies into this sphere; this mechanism might also work for Saturn.

(2) Gas-drag capture during the Solar Nebula period: energy dissipation of planetesimals due to friction with gas from the SN or CPD may transfer an object initially on a hyperbolic orbit onto a closed one \citep{Pollack_1979}. \citet{uk2004} performed a dynamical simulation to show that this mechanism could have worked for the Himalia group of Jupiter and conjectured that a similar mechanism could explain the capture of Phoebe.

(3) Collisions and three-body interactions are also a way of exchanging energy between bodies in a hyperbolic orbit to transfer one of them onto a closed orbit \citep{Colombo_Franklin_1971}. \citet{2007DPS....39.3211N} proposed that irregular moons of giant planets were captured from the planetesimal disk during the encounters between these planets described in the Nice model. They showed via numerical simulations that nearby planetesimals can be deflected into planet-bound orbits during planetary encounters with an efficiency large enough to produce the observed populations of irregular satellites at Saturn, Uranus and Neptune. Jupiter, however, which does not experience encounters with other planets in the Nice model, would have acquired its irregular satellites by a different mechanism. The collision and three-body interaction mechanisms have been explored specifically for Saturn by \citet{Turrini2009}. They solved numerically the inverse capture problem for several of the irregular moons Albiorix and Siarnaq for the prograde group and Mundilfari and Phoebe for the retrograde one, and found that the current population of these moons is compatible with a comet or centaur origin. Based on the analysis of Cassini images of crater impacts at the surface of Phoebe, they also concluded that single-collision capture was possible but of low probability for this moon, thus favoring multiple collision scenarios. 

It remains possible that the most massive irregular moons ($\geq10^{-7}M_{\rm J}$) formed within CPDs \citep{cil21}. This scenario conjectures that they might have been excited by interaction with the other forming satellitesimals prior to settling in the residence domain of irregular moons. However the probability of such an event was found to be low, typically $\sim1\%$ in the case of the Jupiter system \citep{cil21}. 

Close-in exploration of irregular moons systems, still poorly known, by space probes and astrometric studies, revealing their mass, shape, density and surface state, appears to be the only way to discriminate between their different origins and formation scenarios, at Saturn just as at the other giant planet systems.

\subsection{A tentative scenario} \label{tentative}

Our review of the different families of formation scenarios that can be applied to Saturn's moons, illustrated in Fig. \ref{fig:mechanisms}, first confirms that capture is the scenario that applies to irregular moons at Saturn, just as it does at other giant planets. Discussion of the merits and difficulties of the three models proposed for the formation of mid-sized and large moons is more complex. To summarize:

\begin{itemize}
    \item formation by solid re-accretion after a giant impact or the chaotic disruption of a primordial moon system has been proposed in at least two studies. Both studies have difficulties in explaining (1) their current water-rich composition, and (2) their mass-distance relationship. While additional measurements, EOS determinations and dynamical simulations have been identified that could possibly validate a scenario of this family, this first scenario does not appear as the most promising avenue to resolve the puzzle of the formation of the regular moons system;
    \item formation from a CPD, the best candidate for Galilean moons, has a major problem when applied to Saturn. As shown in Fig. \ref{fig:migration}, tracing back in time the current location of mid-size moons to their potential formation site in a primordial CPD takes them inside the Roche limit. The implication is that, if formed in the CPD, they would have been lost into Saturn soon after their accretion. Only Titan could possibly escape this primordial loss, if driven outward fast enough by a tidal resonance locking soon after its accretion;  this negative inference however needs to be taken with due caution, since (1) one can imagine that the Q factor characterizing the efficiency of tidal coupling of moons to Saturn may have evolved in the course of Saturn's life, and (2) that Q also depends on the internal structure, which is still poorly known. 
    \item finally, formation of small and mid-sized moons by viscous expansion of a massive disk appears as a consistent scenario for the following two reasons: (1) the outward edge of Saturn's primordial magnetic cavity (if any) was likely inside the Roche radius; (2) tracing back in time the radial position of these moons (Fig. \ref{fig:migration}) strongly suggests they have been inside or at the edge of the Roche limit at the time of their formation. Of course one should keep in mind that this reconstruction of the moons' past orbital distances that go back to the Roche limit ignores the satellites' mutual interactions, including orbital resonances, for the sake of simplification : it is therefore for the time being a promising theoretical concept, still to be evolved into a detailed model. This scenario is also the one that provides a natural explanation for the mass-distance relationship illustrated in Figs. \ref{fig:architecture} and \ref{fig:Saturnmoons}. 
    {Note that whether the current rings are the remnants of this massive disk \citep{Salmonetal2010,cri19} or a recent addition to the Saturn system \citep{ies19} does not matter here: the massive disk could have disappeared completely before new young rings form \citep{est23}.}
\end{itemize}

Building on these elements, we propose a tentative scenario that can be applied in the same terms to the formation of the Saturnian and Jovian systems, and validate this scenario by a quantitative assessment of the relative magnitudes of the different ``critical radii'' of the problem: planet radius, Roche limit and size of the primordial magnetospheric cavity.

\begin{figure}[htbp]%
\centering
\includegraphics[width=0.7\textwidth]{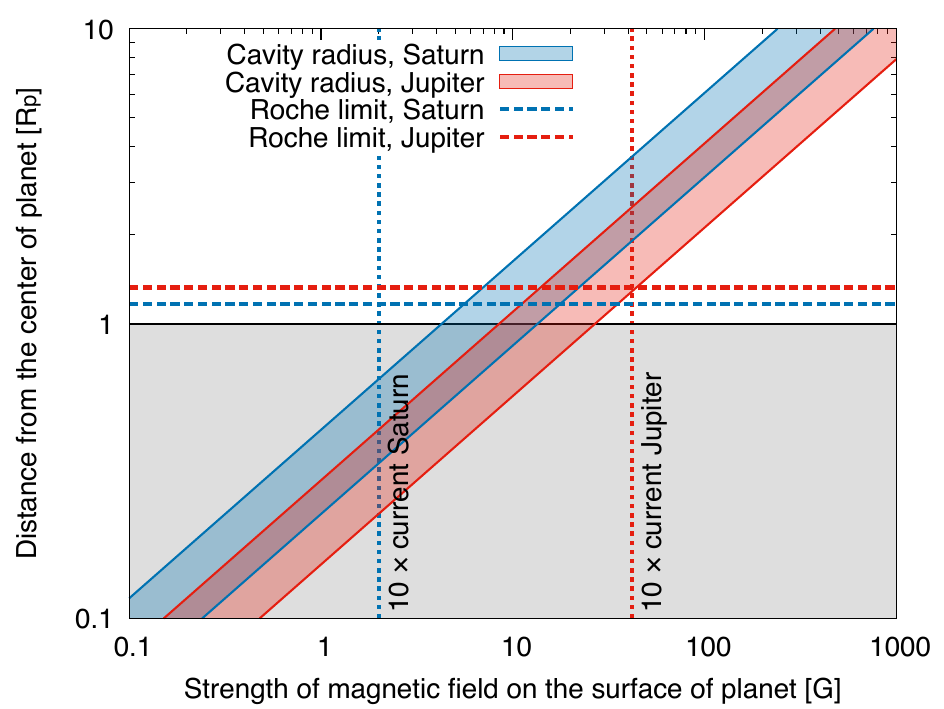}
\caption{Normalized radii of the disk inner cavities truncated by the magnetic fields of young Saturn and Jupiter predicted by Eq. (\ref{Rcav}) (shaded blue and red areas, respectively). We assume gas accretion rates of young Saturn and Jupiter as $\dot{M}_{\rm g}=0.006-0.06$ and $0.02-0.2~M_{\rm J}~{\rm Myr}^{-1}$, respectively, consistent with the typical assumptions based on the gas-starved disk model \citep{can02,shi17,shi19}. The blue and red dashed lines are the Roche limits of young Saturn and Jupiter, respectively, where the planet density is $\rho_{\rm p}=0.20$ and $0.39~{\rm g~cm}^{-3}$ (assuming 1.5 times larger radii than the current \citep{for11}), and the moon density is $\rho_{\rm m}=1.9$ and $2.5~{\rm g~cm}^{-3}$ (Titan and Io; see Fig. \ref{fig:density}). The blue and red vertical dotted lines represent the strengths of the magnetic fields of young Saturn and Jupiter at their surfaces, assumed to be 10 times larger than their current value, $B_{\rm s}=2.0$ and $42~{\rm G}$, respectively \citep{con93,rei10}. The black horizontal line represents the surfaces of the planets (i.e. the planet radii).
}\label{fig:Rcav}
\end{figure}

To set this scenario, let us start from a quantitative evaluation of the truncation radii associated with the magnetospheric cavities of young Saturn and Jupiter. The radius of the inner cavity formed by the pole-aligned dipole magnetic field of the planet is.
\begin{equation}
R_{\rm cav}=\left(\dfrac{\mathcal{M}^{4}}{4GM_{\rm p}\dot{M}_{\rm g}^{2}}\right)^{1/7}
\label{Rcav}
\end{equation}
where $\mathcal{M}=B_{\rm s}R_{\rm p}^{3}$ \citep{tak22}. At the time when gas accretion was ongoing and moons formed in their CPD, planet radii, $R_{\rm p}$, were approximately 1.5 times larger than current ones \citep{for11}. The strength of magnetic fields on the surfaces of planets, $B_{\rm s}$, were also about 10 times larger than the current ones \citep{chr09,rei10}. Figure \ref{fig:Rcav} shows that the predicted truncation radius by the magnetospheric cavity of young Saturn is smaller than the planetary radius when $B_{\rm s}=2.0~{\rm G}$, 10 times larger than the current value \citep{con93}, and $\dot{M}_{\rm g}=0.006-0.06~M_{\rm J}~{\rm Myr}^{-1}$, consistent with the gas-starved disk model \citep{can02}. This means that there was no inner cavity during moon formation and the CPD had a direct connection with the surface of Saturn. In contrast, the predicted truncation radius of young Jupiter is larger than the planetary radius when $B_{\rm s}=42~{\rm G}$ (10 times larger than the current value \citep{con93}) and $\dot{M}_{\rm g}=0.02-0.2~M_{\rm J}~{\rm Myr}^{-1}$ (consistent with the gas-starved disk model and the ice mass fractions of the Galilean moons \citep{can02,shi17,shi19}).

The Roche limits of young Saturn and Jupiter are
\begin{equation}
R_{\rm Roche}=2.456\left(\dfrac{\rho_{\rm p}}{\rho_{\rm m}}\right)^{1/3}R_{\rm p},
\label{RRoche}
\end{equation}
where $\rho_{\rm p}$ and $\rho_{\rm m}$ are the densities of planets and moons, respectively. Figure \ref{fig:Rcav} shows that the Roche limits of both Jupiter and Saturn were larger than the surfaces of the planets, but only the cavity radius of Jupiter was larger than the Roche limit. Note however that the evolution of planetary atmospheres is not well understood, so that the Roche limits could have been smaller than the ``swollen'' planet radii \citep{for11}.

\begin{figure}[htbp]%
\centering
\includegraphics[width=0.9\textwidth]{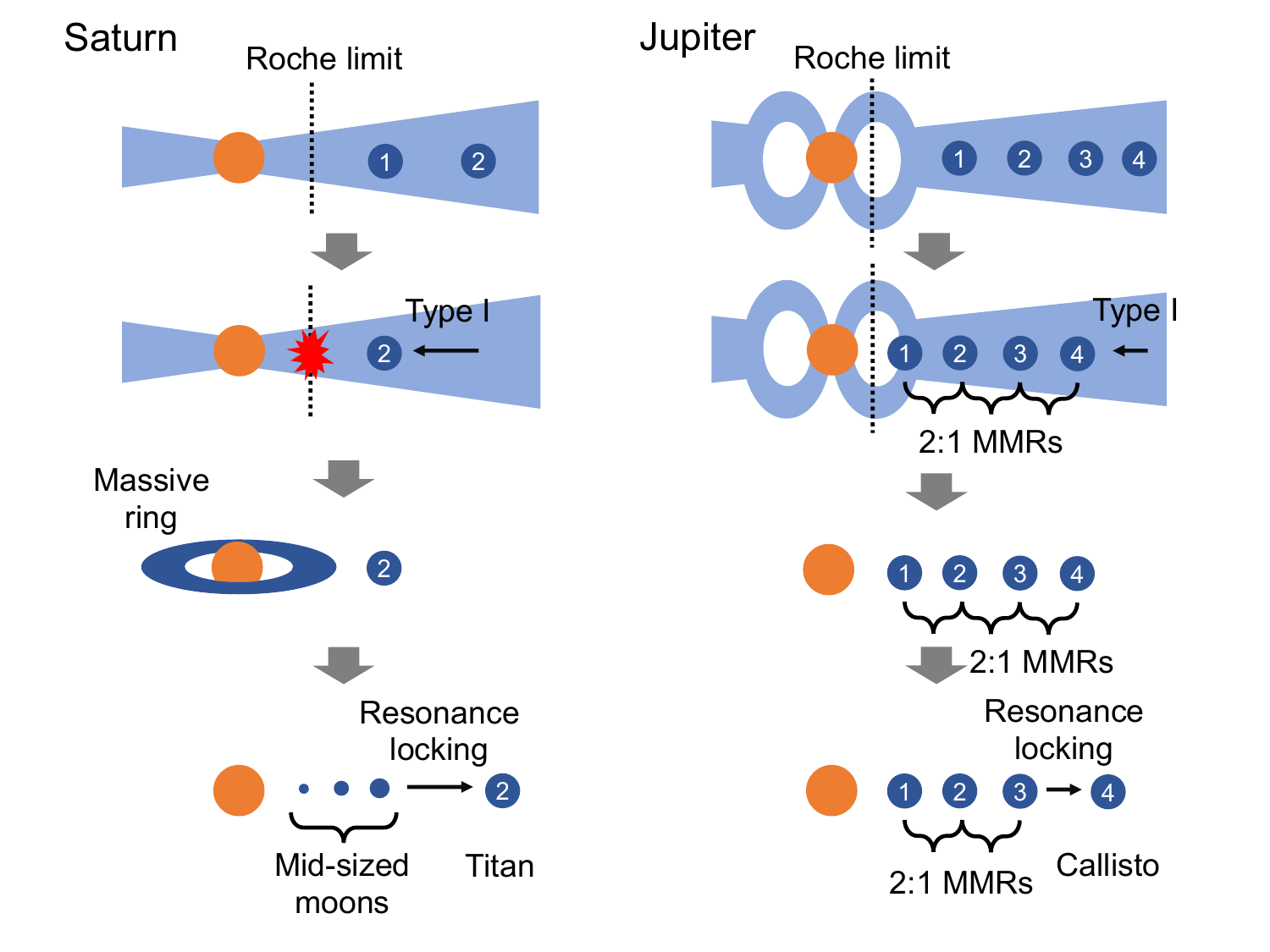}
\caption{Schematic picture of the tentative scenario proposed here for the formation of mid-sized and large moons of Saturn and Jupiter. 
}\label{fig:possible_scenario}
\end{figure}

Based on these estimates, we propose a dual formation scenario for the Saturnian and Jovian systems. The main elements of this scenario are schematically presented in Fig. \ref{fig:possible_scenario}. Let us consider Saturn first, and a situation in which two large moons first form in its CPD. These large moons should have similar masses, according to both the pebble and satellitesimal accretion scenarios, because their mass at formation is determined by the Type I migration timescale or by the pebble isolation mass \citep{canup2006common,ron20}. Let us assume that both moons have Titan's current mass. Similarly to the scenario proposed by \citet{can10}, the inner moon (Moon 1) crosses the Roche limit during its Type I inward migration and its ice-rich mantle (if it is already differentiated) is stripped off, delivering ice fragments to a forming disk, while its rocky core resists the tidal force and falls into Saturn. Then, following dissipation of the gas disk, a massive icy ring forms out of the fragments of Moon 1, as predicted by the SPH simulation in \citet{can10}. Mid-sized moons could then form by viscous expansion of this massive ring, as described in Section \ref{massivedisk} above. A constraint placed on this scenario by the current silicate content of Saturn's mid-size moons (i.e., Figure~\ref{fig:density}) is that the primordial rings born out of the stripping off of Moon 1 should contain enough silicates to provide them to these moons - or that the current rings, with their high ice content, were formed later after erosion of a primordial ring. This connects us to the issue of the age of the rings, addressed by Crida et al. (2024) in this topical collection.

At the same time, the outer moon (Moon 2) could be at the origin of Titan, but for this to be possible it must have halted its inward migration during the gas disk phase before falling inside the Roche limit. This requires that Moon 2 formed late, just before the CPD dissipated, so that its migration was slower and short. Alternatively, powerful resonance locking tides from Saturn may have captured Moon 2 and pushed it outwards as it was migrating in due to the CPD. Keeping Moon 2 is a bit tricky, but in any case the fast migration of Titan observed by \citet{lai20} allows for more inwards migration of this Moon 2 than in previous models.

While Iapetus is not directly addressed in this scenario, it leaves room for Iapetus to have been able to form at a larger distance in the same CPD that gave birth to moons 1 and 2, as a third object outside the orbit of Moon 2. Its small mass would have avoided its significant Type I migration from the time of its origin.

A strong point of this scenario is that it also matches well the formation of Jovian moons: first, the four large Galilean moons form in the CPD by pebble or satellitesimal accretion. Then they move inward by Type I migration, and the innermost one (Moon 1) stops at the inner cavity, which resides outside the Roche limit because of the stronger magnetic field of Jupiter \citep{shi19}. Then the other three moons continue to drift inward until they are captured into the 2:1 MMR one by one, forming the Laplace resonance, just like giant planets around stars in low viscosity disks \citep{gri23}). After dissipation of the CPD gas, the outermost moon (Moon 4, i.e., Callisto) may have left the resonance and moved outward under the effect of resonance locking, as also suggested by orbital dynamical simulations \citep{lar23}.

\section{Conclusions and directions for future work}\label{summary}

In this article, we explored the different formation scenarios of the Kronian moons system in the context of a highly dissipative Saturn, with the objective of identifying at least one tentative scenario consistent with both observations and current theories and models.
In order to do so, we first reviewed the diversity of moons and moon system architectures existing among the four giant planet systems of our Solar System, in the light of past and current Earth-based telescopic and space mission observations. This allowed us to identify the specific features of the different categories of Saturnian moons (small, mid-sized and large) that have to be explained by candidate formation scenarios.

We then reviewed the different types of moon formation scenarios that have been proposed for the four giant planet systems: result of a giant impact; formation by accretion in a circumplanetary disk (CPD); formation by viscous expansion from a massive disk; capture of small bodies from Sun-centered orbits. For each category of formation scenario, we discussed its relevance to the formation of Saturnian moons and of the Saturn system as a whole. We discussed the merits and difficulties of the different formation scenarios on the basis of available observations and in the context of the observed fast migration of moons. Finally, we proposed one tentative formation scenario that can be applied in the same terms to the cases of Saturn and Jupiter, starting from the assumptions that moons formed inside CPDs and migrated inwards and that the strength of the magnetic fields, and thus the radius of the inner magnetic cavity, was different between the two planets.

This scenario, like alternative ones, will need further observations, theories and simulations to be fully validated: all CPD-based scenarios strongly depend on knowledge of the structure of the CPD, which is unfortunately still controversial. There will be two ways to investigate CPDs in more detail in the near future: 1) numerical simulations with higher resolution and 2) observations of CPDs around young exoplanets. 

Current numerical simulations still do not have enough spatial resolution to reveal the detailed gas structures and flows from inside the satellite formation region (inner than 0.01 Hill radius) to the entire CPDs. High resolution MHD simulations including the whole process of gas accretion on planets are also necessary. 

Observations of CPDs around forming exoplanets will be another strong contribution to revealing the conditions prevailing in CPDs. There has already been one detection of a CPD around a gas-accreting exoplanet: dust continuum at (sub)millimeter wavelength with ALMA from PDS~70~c \citep{ise19,ben21}. A few candidate CPDs around planets embedded in gas gaps have also been identified by their signature in the dust continuum and in the gas velocity distortion \citep{and21,bae22}. Such observations can constrain the properties of CPDs as well as those of their host planets \citep{shi24}. As the number of observations will progressively increase, statistical information will also progressively become available.

Progress in our knowledge of the interiors of giant planets, as recently accomplished for Jupiter with the Radio Science experiment on board NASA's Juno, is also greatly needed to better quantify the efficiency of tidal coupling between moons and Jupiter (the Q factor) and the actual rate of tidal migration of moons. In the framework of this Topical Series dedicated to "a highly dissipative Saturn", we did not explore all the possible complexities related to the Q factor. No doubt, this will be an important avenue for future research, heavily dependent on improved knowledge of Saturn's and more generally giant planets interiors. 

Finally, new and additional information on the icy material content of giant planets regular moons will also provide still missing ground truth for formation scenarios. Detailed analysis of the cratering records and related ages of moons surfaces, recently reviewed in detail by \citet{Bottke2024}), provides important constraints on the different formation scenarios, in particular by favoring formation of mid-size moons very early in the History of the Solar System, provided that most of the colliders were in heliocentric orbit and given we know the primordial population of impactors. In addition, determining the chemical and isotopic compositions of the surfaces and plumes of icy moons will provide critical information on the thermal history of the reservoirs of icy materials which have existed in protoplanetary and circumplanetary disks \citep{hor08,bie20}. There is no doubt that ESA's JUICE mission, as well as NASA's Europa clipper mission, which are both heading to the icy moons of Jupiter, will return to Earth critical new observations of the plumes and atmospheres of the Jovian icy moons in the next decade, likely to be followed by missions to Titan and Enceladus in the Saturn system, as described in the Horizon 2061 Planetary Exploration foresight exercise \citep{bla22}. This new wave of missions to the gas giants of our solar system is our best chance to far better constrain the formation scenarios of their moons in the coming decades.

\backmatter


\bmhead{Acknowledgments}
We thank William F. Bottke and the other anonymous reviewer for the very valuable comments. This paper resulted from the Workshop ``New Vision of the Saturnian System in the Context of a Highly Dissipative Saturn'' held at the International Space Science Institute (ISSI) in Bern (Switzerland).

\bmhead{Author Contribution}
All authors have contributed to this review article.

\bmhead{Funding}
Y.S. contributed to this work within the framework of the NCCR PlanetS supported by the Swiss National Science Foundation under grants 51NF40\_182901 and 51NF40\_205606, and acknowledges support from the Swiss National Science Foundation under grant 200021\_204847 `Planets In Time'. This work was supported by JSPS KAKENHI Grant Numbers JP22H01274, JP23K22545, and JP24K22907.

\bmhead{Data Availability}
Not applicable.

\bmhead{Code Availability}
Not applicable.

\section*{Declarations}
\bmhead{Competing Interests}
The authors have no relevant financial or non-financial interests to disclose.

\bmhead{Open Access}
This article is licensed under a Creative Commons Attribution 4.0 International License, which permits use, sharing, adaptation, distribution and reproduction in any medium or format, as long as you give appropriate credit to the original author(s) and the source, provide a link to the Creative Commons licence, and indicate if changes were made. The images or other third party material in this article are included in the article's Creative Commons licence, unless indicated otherwise in a credit line to the material. If material is not included in the article's Creative Commons licence and your intended use is not permitted by statutory regulation or exceeds the permitted use, you will need to obtain permission directly from the copyright holder. To view a copy of this licence, visit \url{http://creativecommons.org/licenses/by/4.0/}.

\bibliography{chapter12}


\end{document}